\documentclass[manuscript]{acmart}
\AtBeginDocument{%
  \providecommand\BibTeX{{%
    \normalfont B\kern-0.5em{\scshape i\kern-0.25em b}\kern-0.8em\TeX}}}

\makeatletter
\setcopyright{none}
\makeatother
\copyrightyear{2024}
\acmYear{2024}
\acmDOI{}

\usepackage{tabularx}
\usepackage{adjustbox}
\usepackage{longtable}
\usepackage{rotating}
\usepackage{subcaption}
\usepackage{tikz}
\usepackage{caption}

\begin{document}

%%
%% The "title" command has an optional parameter,
%% allowing the author to define a "short title" to be used in page headers.
\title{Assistive XR research for disability at ACM ASSETS: A Scoping Review }

%%
%% The "author" command and its associated commands are used to define
%% the authors and their affiliations.
%% Of note is the shared affiliation of the first two authors, and the
%% "authornote" and "authornotemark" commands
%% used to denote shared contribution to the research.
\author{Puneet Jain}
\affiliation{%
  \institution{Concordia University}
 % \streetaddress{1 Th{\o}rv{\"a}ld Circle}
  \city{Montreal}
  \country{Canada}}
\email{puneet798@gmail.com}
%%
%% By default, the full list of authors will be used in the page
%% headers. Often, this list is too long, and will overlap
%% other information printed in the page headers. This command allows
%% the author to define a more concise list
%% of authors' names for this purpose.
%\renewcommand{\shortauthors}{Trovato and Tobin, et al.}

%%
%% The abstract is a short summary of the work to be presented in the
%% article.

\begin{abstract}
Despite the rise in affordable eXtended Reality (XR) technologies, accessibility still remains a key concern, often excluding people with disabilities from accessing these immersive XR platforms. Consequently, there has been a notable surge in HCI research on creating accessible XR solutions (also known as, assistive XR). This increased focus in assistive XR research is also reflected in the number of research and innovative solutions submitted  at the ACM Conference on Accessible Computing (ASSETS), with an aim to make XR experiences inclusive for disabled communities. However, till date, there is little to no work that provides a comprehensive overview of state-of-the-art research in assistive XR for disability at ACM ASSETS, a premier conference dedicated for research in HCI for people with disabilities.

This study aims to fill this research gap by conducting a scoping review of literature delineating the key focus areas, research methods, statistical and temporal trends in XR research for disability at ACM ASSETS (2019-2023). From a pool of 1595 articles submitted to ASSETS, 26 articles are identified that specifically focus on XR research for disability. Through a detailed analysis, 6 key focus areas of XR research explored at ACM ASSETS are identified and a detailed examination of each is provided. Additionally, an overview of multiple research methods employed for XR research at ASSETS is also presented. Lastly, this work reports on the statistics and temporal trends regarding the number of publications, XR technologies used, disabilities addressed, and methodologies adopted for assistive XR research at ASSETS, highlighting emerging trends and possible future research directions.
\end{abstract}


%%
%% Keywords. The author(s) should pick words that accurately describe
%% the work being presented. Separate the keywords with commas.

%% A "teaser" image appears between the author and affiliation
%% information and the body of the document, and typically spans the
%% page.

%\received{20 February 2007}
%\received[revised]{12 March 2009}
%\received[accepted]{5 June 2009}

%%
%% This command processes the author and affiliation and title
%% information and builds the first part of the formatted document.
\maketitle

\section{Introduction}

With the recent proposal of building a new XR social-media platform, \textit{The Metaverse}, by Meta (formerly Facebook) for its 3.2 billion users, there has been an increasing interest in the emerging technologies of eXtended Reality. XR (or eXtended Reality) is an umbrella term for computer-generated environments (Virtual/Augmented/Mixed Reality (VR/AR/MR)) experienced through body-centric technologies (e.g. head-mounted displays, hand-held controllers, mobile-displays) that bridge the physical body within a continuum of real and virtual space \cite{milgram1994taxonomy}. However, the demanding head-movements (360-degree point of view), use of "hand-held" controllers to interact with these devices, and advertising of VR/AR as "vision-centered" devices excludes people with disabilities (e.g. sensorimotor impairment, people who are blind) from using these devices. Hence, revealing the implicit assumptions about human bodies and their sensory capabilities imbibed in the design of these XR technologies,  what Gerling and Spiel refer to as the `corporeal standard' (i.e, a non-disabled body) \cite{gerling2021critical}. Therefore, there has been an increased focus of research in HCI on "assistive XR", that is, creating accessible XR solutions for people with disabilities (\cite{biswas2021adaptive}; \cite{creed2023inclusive}; \cite{sidartoaccessibility}). Meanwhile, ACM Conference on Accessible Computing (ASSETS), a premier conference dedicated for research in HCI for people with disabilities, has also witnessed growing research and innovative solutions, addressing accessibility of XR technologies for people with varied disabilities. A recent literature review by Mack et al. \cite{mack2021we} provided a detailed analysis on the historical and current trends in "accessibility research" at ASSETS conference over 10 year period. While the review by the authors provided an extensive analysis on the use of accessible digital technologies (including XR), their goal differed from providing a focused overview on XR research at ASSETS. Hence, the work on assistive XR at ASSETS still misses a structured overview that provides: 1) A coded analysis that identifies the focal areas of XR research at ASSETS, 2) Research methods adopted by the HCI community for XR research for disabled communities, and 3) Temporal trends and possible future directions in the research published at ASSETS conference at the intersection of assistive XR and disability. 

This work aims to address this research gap by scoping the research at the intersection of VR/AR/MR technology built for (and with) people with disabilities at ACM ASSETS conferences in the time range of 2019-2023. This range of time is chosen specifically to focus on the state of the art research in assistive XR following the expansion of interest in XR by leading corporations exemplified by Meta's claim of building `The Metaverse' in 2021, followed by Apple's launch of Vision Pro headsets in 2024 - advertising them as the technologies for `new era of "spatial" computing'. Additionally, founding of community driven symposiums such as XR Access \cite{xraccess} in 2019 to make "XR inclusive for all" strengthens the need for such a scoping review that brings out the state of the art and future trends in XR research at ACM ASSETS conference (known for presenting premier research in computing for accessibility). Concretely, this work presents a scoping review of 26 research papers published at ACM ASSETS (from 2019-2023) that specifically focused on assisitve XR solutions for people with disabilities (refer Tables \ref{tab:mytable1}, \ref{tab:mytable2}, and \ref{tab:mytable3}). These 26 publications were  searched, identified, and screened from 1595 publications published at the ASSETS conference in the time-range of 2019-2023 using a set of keywords that cover the broad spectrum of XR research for disability (refer Figure \ref{PRISMAflowchart}). Finally, using qualitative coding, data was extracted and analysed to address the following research questions: 

\begin{itemize}
    \item[\textbf{RQ1:}] What are the key focus areas addressed at ASSETS conference at the intersection of XR technology, accessibility, and disability?
    \item[\textbf{RQ2:}] Which prevalent research methods have been adopted by researchers for building assistive XR technologies for (and with) people with disabilities?
    \item[\textbf{RQ3:}] What are the general statistical and temporal trends in assistive XR research for disability at ACM ASSETS over the period 2019-2023?
\end{itemize}

Addressing these research questions, the following sections firstly, provide some background and related work done in XR research for disability (refer Section \ref{Background}). In Section \ref{Methodologies}, the methodologies to conduct this scoping review are described, further outlining the codebook created for the analysis. This section also lists out all the 26 identified papers categorized according to the identified codes in the order of their publications. Section \ref{Results} shares the results from the data extraction and analysis of the 26 identified papers, outlining the 6 key focus areas (with statistics) that are being explored in assisitve XR research at ASSETS for disability (addressing \textbf{RQ1}). Furthermore, the results address the \textbf{RQ2} by providing an overview of the research methods used by researchers publishing at ACM ASSETS in XR research for disabled communities. Lastly, reflecting on \textbf{RQ3}, the results also provide statistics and temporal trends of the number of publications published, XR technologies used, disabilities addressed, and methodologies employed over the period 2019-2023 at ASSETS for research in assisitve XR. The paper ends with a conclusion, limitations and future work.

\section{Background}\label{Background}

The use of eXtended Reality (XR) technologies in context of disability is not a new area of research. For instance, VR has been widely used to simulate environments that can enable people with disabilities to "\textit{engage in a range of activities in a simulator relatively free from the limitations imposed by their disability}" \cite{wilson1997virtual}. Hence, offering disabled communities therapeutic and rehabilitative solutions ranging from posture and balance control \cite{kim1999new} \cite{sveistrup2004motor} \cite{fung2004locomotor}, aid in navigation \cite{thevin2020x}, stress-reducing simulations \cite{seol2017drop}, to the
treatment of eye conditions (such as amblyopia or “lazy eye”) \cite{seevividly}. Moreover, reference of VR as "ultimate empathy machines" \cite{ventura2020virtual} \cite{hassan2020digitality}, as the technologies that can enable one "\textit{to step into others' shoes}" through simulations, has popularized these technologies as tools that can foster positive attitudes (of able-bodied people) towards disabled communities \cite{chowdhury2019vr}. For example, Pivik et al. \cite{pivik2002using} created a VR program that teaches kids about accessibility and attitudinal barriers encountered by their peers with mobility impairments. Gerling et al. \cite{gerling2014effects} also explored how building persuasive games in VR can creative positive attitudes of the public towards people with disabilities who use wheelchairs. 
Meanwhile, trending away from the use of VR to "fix" disabled people using technology \cite{shew2020ableism} or creating VR simulations to re-construct disability for able-bodied people, HCI research is witnessing an increasing trend to enhance inclusivity in XR by enabling people with disabilities experience and use XR technologies. For instance, Siu et al. \cite{siu2020virtual} created a haptic and auditory
white cane for people with visual impairments enabling them to navigate in complex virtual environments in VR. Gerling et al. \cite{gerling2020virtual} developed three full-body VR game prototypes enabling people using wheelchairs to experience embodied VR games. Wentzel \cite{wentzel2023bring} explored context-aware multi-modal adaptation techniques in VR for people with limited mobilities, creating more accessible VR experiences.

Following such interests in inclusive XR and to investigate the scope of XR research for disability, some researchers conducted scoping reviews on the existent literature as well. Though less in numbers, these scoping reviews have explored in what context XR technologies have been in use for (and by) disabled communities. For instance, Maran et al. \cite{maran2022use} in their scoping study, surveyed how XR technologies are used by and for people with moderate to severe intellectual disabilities (IDs). The authors reported that the majority of XR interventions for people with IDs focused on improving navigation and daily living skills, with an aim to increase autonomy and independence. Similarly, Nabour et al. \cite{nabors2020scoping} conducted a scoping review to investigate VR interventions for people with intellectual disabilities. The researchers summarized in their results that their survey indicates positive outcomes in use of VR for people with intellectual disabilities (IDs) such as use of VR for improving skills and physical conditioning of people with IDs. A recent scoping review published at ACM ASSETS by Li et al. \cite{li2022scoping} reported how Head-Mounted Displays (HMDs) are used for assistance and therapy for people who have visual impairments. Through a structured screening process, the researchers analyzed 61 research articles that employed the use of HMDs for enhancing the visual sense of people with visual impairments. While the researchers expanded previous explorations of scoping reviews (limited to VR) by investigating use of both VR and AR in context of assistance and therapy, the review focused on a particular disability, that is, visual impairments. With ACM ASSETS as one of the premier conferences that focuses on HCI research in context of disability, there is still no scoping review that provides an overview of the state of the art research in assistive XR at ACM ASSETS, detailing the key focus areas, the methodologies used, and trends in assisitve XR research for varied disabilities. This work, thus aims to address this gap by conducting a scoping review in assisitve XR research for disability published at ACM ASSETS for the years 2019-2023. The methodologies to carry out the scoping review are discussed in the next section. 

\section{Methodology}\label{Methodologies}

\begin{figure}[htbp]
  \centering
  \includegraphics[width=0.8\textwidth]{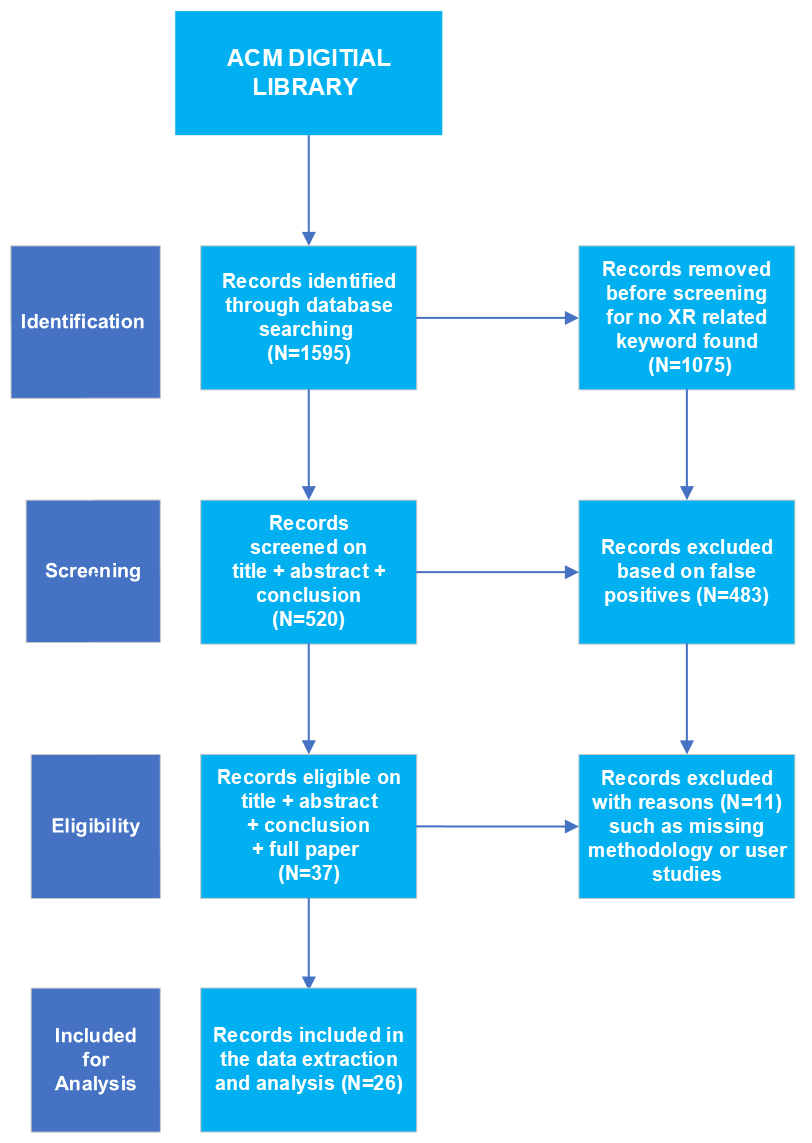}
  \caption{Review Process}
  \label{PRISMAflowchart}
  \Description{Flowchart outlining the scoping review process, including identification, screening, eligibility, and analysis stages, with numbers indicating the quantity of records at each stage.}
\end{figure}

The aim of this paper is to identify the state of the art research in XR for people with disabilities at ASSETS. Hence, a scoping review is conducted with the motivation to “\textit{provide a preliminary assessment of the potential size and scope of available research literature}” \cite{sutton2019meeting} at ASSETS (with N=26) in the time-range 2019-2023. Following the scoping study frame work of  Arksey and O’Malley \cite{arksey2005scoping} and the review procedure and its results using best-practice items recommended by the latest version of the Preferred Reporting Items for Systematic Reviews and Meta-Analyses extension for Scoping Reviews (PRISMA-ScR, 2018) \cite{tricco2018prisma}, the following sections  describe in detail the searching strategy, the screening and elimination process, and coding analysis to identify the major focal themes in context of assistive XR research at ASSETS. See Figure \ref{PRISMAflowchart} for overall review process.

\subsection{Searching Strategy}

ACM digital library was used as the prime source to search all the relevant published articles at ACM ASSETS. Firstly, the author used keywords such as "extended reality" OR "mixed reality" OR "virtual reality" OR "augmented reality" to randomly select 25 publications that focused their research on emerging technologies of VR/AR/MR/XR (selecting 5 publications per year from 2019-2023). The author manually read all the selected 25 publications to identify more keywords (other than those listed above) to be added to the final list of search keywords (to search the publications to be screened later). Since ASSETS conference is dedicated towards accessibility research, hence, search keywords on disability were not considered to avoid redundant addition of articles. It was realized that some publications didn't use the words such as Virtual or Augmented Reality to describe XR technologies but instead used words like "head-mounted displays", "see-through displays" etc. 
Hence, such keywords were also added to the search keywords list to search out all the publications that addressed XR for disability. Overall list of search keywords are shown in Table \ref{searchkeywords_table}. 

\begin{table}[htbp]
  \centering
   \caption{List of search keywords for finding articles published at ASSETS that address XR for disability}
  \label{searchkeywords_table}
  \begin{tabularx}{\textwidth}{|p{0.15\linewidth}|>{\centering\arraybackslash}X|>{\raggedleft\arraybackslash}X|}
    \hline
    Search Keywords & "extended reality" OR "mixed reality" OR "virtual reality" OR "augmented reality" OR "HMD" OR "head-mounted display" OR "virtual environments" OR "see-through displays" \\
    \hline
  \end{tabularx}
\end{table}

\subsection{Screening Strategy}

Using the final list of search keywords, a total of 520 papers were selected from 1595 publications at ASSETS from 2019-2023. Hence, 1075 publications were excluded from the screening process since they did not include any of the search keywords as listed in Table \ref{searchkeywords_table}. To further refine the search, the abstract, title and the conclusions of the 520 selected publications were manually read by the author to exclude the papers which were not relevant, for example, did not use XR technologies in their research (or partially referenced XR research). For instance, many publications (from the selected 520 articles) referenced VR/AR works in the "related works" section but did not conduct research on XR (that is, false positives). Hence, from the 520 selected articles, 483 publication were excluded for the next stage of analysis. While all publications including the full articles, short papers, demonstration, posters, and extended abstracts were included in the screening process, some of them were removed from the list if they did not qualify the criteria (decided by the author) such as: not conducting user studies or providing specific details of the user studies (i.e. what kind of qualitative/quantitative/mixed method approach was adopted for research) - resulting in a total of 26 papers eligible for the analysis (please refer to Figure \ref{PRISMAflowchart}). 

\subsection{Qualitative coding and code book}

Firstly, to identify a high-level categorization in state of the art research in XR over the past 5 year period at ASSETS, the author charted the data in 6 main categories: "Year of publication", "Disability addressed" (e.g. vision-impairment, deaf or hard of hearing),  "XR technology used" (e.g. VR or AR), "Objective of research" (aid in navigation in daily life using VR), "Methodology used for study" (e.g. qualitative, quantitative or mixed-methods etc.) and the "Targeted end users" (e.g designers, people with disabilities, therapists etc.) for conducting the research. Secondly, using the charted data a codebook was generated based on the identified categories  (along with the subcodes).

\begin{table}[htbp]
  \centering
  \caption{Generated codebook for analysis}
  \label{codebook}
  \begin{tabularx}{\textwidth}{|p{0.2\linewidth}|X|}
    \hline
    Category & Codes (subcodes)  \\
    \hline 
    Disability addressed & Vision Impairment (Blind, Low vision, Amblyopia); Deaf or Hard of hearing; Invisible disabilities (cognitive disabilities); Sensorimotor impairment (upper body, limited mobility); Autism; Aphasia; All disabilities \\ 
    \hline 
    XR technology used & Virtual Reality; Augmented Reality (phone-based, smart glasses) \\
    \hline
    Objectives of research  & Aid in Navigation; Aid in Body-orientation; Interaction modalities; Therapy and Rehabilitation; Enhancing social inclusion and engagement in XR; Enhancing communication and language support; Using XR to support other assisitve technologies \\
    \hline
    Methods used for study & Qualitative; Quantitative; Mixed-methods \\
    \hline
    Targeted end users & Designers; Disabled groups; Therapists (Occupational therapists); Educational professionals; Developers \\
    \hline    
  \end{tabularx}
\end{table}

Table \ref{codebook} shows the generated code-book. Table \ref{tab:mytable1}, \ref{tab:mytable2}, and \ref{tab:mytable3} detail out the list of all the selected 26 publications, codified on the identified codes in the order of their publication years. Furthermore, addressing \textbf{RQ1}, the process of iterative search/analysis (using the generated code "Objectives of research") from the 26 selected publications, enabled the author to point out main areas of focus in XR research for disability at ASSETS (2019-2023). Hence, the identified key focus areas are as follows: \textbf{a)} XR for Navigation, Orientation, and Interaction, \textbf{b)} XR for therapeutic and rehabilitation research, \textbf{c)} XR to enhance social inclusion and engagement for people with disabilities, \textbf{d)} XR for Communication and Language Support, \textbf{e)} General Accessibility in XR, and \textbf{f)} XR for supporting research in other assisitve technologies. These key focus areas are described in detail with statistics in Section \ref{Results}. 

\subsection{Limitations and Potential Bias}

The author acknowledges that there could be a potential bias in the search strategy with the keywords selected. Since this work has been manually carried out by a single author, the quantity of papers searched for finalizing the keywords was low, that is, 25 randomly selected papers (5 in each year from 2019-2023 that had "extended reality" OR "mixed reality" OR "virtual reality" OR "augmented reality" in their title) from a range of 1595 papers submitted at ASSETS during 2019-2023. In addition, limiting the search to just ACM digital library might have resulted in missing certain publications (however, is unlikely).

\begin{table}[htbp]
  \centering
  \begin{adjustbox}{angle=90}
    \begin{tabularx}{\textheight}{|p{0.35\linewidth}|>{\centering\arraybackslash}X|>{\centering\arraybackslash}X|>{\centering\arraybackslash}X|>{\centering\arraybackslash}X|>{\centering\arraybackslash}X|>{\centering\arraybackslash}X|}
      \hline
      Title & Year of publication & Disability addressed & XR technology used  & Objective(s) of research & Method(s) used for study & Targeted end users \\
      \hline
      Identifying Comfort Areas in 3D Space for Persons with Upper Extremity Mobility Impairments Using Virtual Reality & 2019 & Sensorimotor disability (Upper impairments) & VR & XR for supporting research in other assisitve technologies & Quantitative & Occupational therapists and Designers \\
      \hline
      Leveraging Augmented Reality to Create Apps for People with Visual Disabilities: A Case Study in Indoor Navigation & 2019 & Vision impairments &  AR (Phone-based) & XR for Navigation, Orientation, and Interaction & Mixed-Methods & Disabled groups \\
      \hline 
      Insights for More Usable VR for People with Amblyopia & 2019 & Vision impairment (Amblyopia) & VR & XR for Navigation, Orientation, and Interaction & Qualitative & Designers \\
      \hline
      CLEVR: A Customizable Interactive Learning Environment for Users with Low Vision in Virtual Reality & 2020 & Vision impairments (Low Vision) & VR & XR for Navigation, Orientation, and Interaction & Qualitative & Disabled groups \\
      \hline
      Teaching ASL Signs using Signing Avatars and Immersive Learning in Virtual Reality & 2020 & Deaf or Hard of Hearing & VR & XR for Communication and Language Support & Qualitative & Disabled groups \\
      \hline
      “I just went into it assuming that I wouldn't be able to have the full experience”: Understanding the Accessibility of Virtual Reality for People with Limited Mobility & 2020 & Sensorimotor Impairments & VR & Accessibility in XR & Qualitative & Designers \\
      \hline
      Making Mobile Augmented Reality Applications Accessible & 2020 & Vision impairments & AR (Phone-based) & Accessibility in XR & Qualitative & Developers and Designers \\
      \hline
      AIGuide: An Augmented Reality Hand Guidance Application for People with Visual Impairments & 2020 & Vision Impairments & AR (Phone-based) & XR for Navigation, Orientation, and Interaction & Mixed Methods & Disabled groups \\
      \hline 
      An Exploratory Study on Supporting Persons with Aphasia in Pakistan: Challenges and Opportunities & 2020 & Aphasia & VR & XR for therapeutic and rehabilitation research & Qualitative & Disabled groups \\   
      \hline
      Chat in the Hat: A Portable Interpreter for Sign Language Users & 2020 & Deaf or Hard of Hearing &  AR (Vuzix Blade smartglasses) & XR for Communication and Language Support & Mixed-Methods & Disabled groups \\
      \hline
    \end{tabularx}
  \end{adjustbox}
  \caption{The identified 26 papers with corresponding categories}
  \label{tab:mytable1}
\end{table}

\begin{table}[htbp]
  \centering
  \begin{adjustbox}{angle=90}
    \begin{tabularx}{\textheight}{|p{0.35\linewidth}|>{\centering\arraybackslash}X|>{\centering\arraybackslash}X|>{\centering\arraybackslash}X|>{\centering\arraybackslash}X|>{\centering\arraybackslash}X|>{\centering\arraybackslash}X|}
      \hline
      Title & Year of publication & Disability addressed & XR technology used  & Objective(s) of research & Method(s) used for study & Targeted end users \\
      \hline
      Nearmi: A Framework for Designing Point of Interest Techniques for VR Users with Limited Mobility & 2021 & Sensorimotor impairments & VR & XR for Navigation, Orientation, and Interaction & Qualitative & Designers \\
      \hline 
      Investigating Sign Language Interpreter Rendering and Guiding Methods in Virtual Reality 360-Degree Content & 2022 & Deaf or Hard of Hearing &  VR & XR for Navigation, Orientation, and Interaction & Qualitative & Disabled groups \\
      \hline 
      VRBubble: Enhancing Peripheral Awareness of Avatars for People with Visual Impairments in Social Virtual Reality & 2022 & Vision impairments & VR & XR for Navigation, Orientation, and Interaction & Mixed-Methods & Disabled groups \\
      \hline 
      SoundVizVR: Sound Indicators for Accessible Sounds in Virtual Reality for Deaf or Hard-of-Hearing Users & 2022 & Deaf or Hard of Hearing & VR & XR for Navigation, Orientation, and Interaction & Mixed-Methods & Disabled groups \\
      \hline 
      College Students’ and Campus Counselors’ Attitudes Toward Teletherapy and Adopting Virtual Reality (Preliminary Exploration) for Campus Counseling Services & 2022 & Sensorimotor Impairments and cognitive disabilities & VR & XR for therapeutic and rehabilitation research & Qualitative & Disabled groups \\
      \hline
      “It’s Just Part of Me:” Understanding Avatar Diversity and Self-presentation of People with Disabilities in Social Virtual Reality & 2022 & All & VR & XR to enhance social inclusion and engagement for people with disabilities & Qualitative & Designers \\
      \hline
      Access on Demand: Real-time, Multi-modal Accessibility for the Deaf and Hard-of-Hearing based on Augmented Reality & 2022 & Deaf or Hard of hearing & AR (Vuzix Blade AR smart glasses) & XR for Communication and Language Support & Qualitative & Disabled groups \\
        \hline 
      Designing a Customizable Picture-Based Augmented Reality Application For Therapists and Educational Professionals Working in Autistic Contexts & 2022 & Autism & AR (phone-based) & XR for therapeutic and rehabilitation research & Qualitative & Therapists and Educational professionals \\
      \hline
    \end{tabularx}
  \end{adjustbox}
  \caption{The identified 26 papers with corresponding categories}
  \label{tab:mytable2}
\end{table}

\begin{table}[htbp]
  \centering
  \begin{adjustbox}{angle=90}
    \begin{tabularx}{\textheight}{|p{0.35\linewidth}|>{\centering\arraybackslash}X|>{\centering\arraybackslash}X|>{\centering\arraybackslash}X|>{\centering\arraybackslash}X|>{\centering\arraybackslash}X|>{\centering\arraybackslash}X|}
      \hline
      Title & Year of publication & Disability addressed & XR technology used  & Objective(s) of research & Method(s) used for study & Targeted end users \\
      \hline 
      Comparing Locomotion Techniques in Virtual Reality for People with Upper-Body Motor Impairments & 2023 & Sensorimotor Impairments & VR & XR for Navigation, Orientation, and Interaction & Mixed-Methods & Designers \\
      \hline 
      Embodied Exploration: Facilitating Remote Accessibility Assessment for Wheelchair Users with Virtual Reality & 2023 & Wheelchair users & VR & XR for Navigation, Orientation, and Interaction & Qualitative & Disabled groups \\
      \hline 
      A Diary Study in Social Virtual Reality: Impact of Avatars with Disability Signifiers on the Social Experiences of People with Disabilities & 2023 & All & VR & XR to enhance social inclusion and engagement for people with disabilities & Qualitative & Designers \\
      \hline 
      The Eyes Have It: Visual Feedback Methods to Make Walking in Immersive Virtual Reality More Accessible for People With Mobility Impairments While Utilizing Head-Mounted Displays & 2023 & Sensorimotor impairments & VR & XR for Navigation, Orientation, and Interaction & Mixed-Methods & Disabled groups \\
      \hline 
      “The Guide Has Your Back”: Exploring How Sighted Guides Can Enhance Accessibility in Social Virtual Reality for Blind and Low Vision People & 2023 & Vision impairments & VR & XR for Navigation, Orientation, and Interaction & Qualitative & Disabled groups \\
      \hline 
      Investigation into Stress Triggers in Autistic Adults for the Development of Technological Self-Interventions & 2023 & Autism & VR and AR & XR for therapeutic and rehabilitation research & Qualitative & Disabled groups \\
      \hline 
      “Invisible Illness Is No Longer Invisible”: Making Social VR Avatars More Inclusive for Invisible Disability Representation & 2023 & Invisible Disabilities & VR and AR & XR to enhance social inclusion and engagement for people with disabilities & Qualitative & Disabled groups \\
      \hline 
      A Demonstration of RASSAR: Room Accessibility and Safety Scanning in Augmented Reality & 2023 & Sensorimotor impairments and Vision Impairment & AR (phone-based) & XR for Navigation, Orientation, and Interaction & Qualitative & Disabled groups \\
      \hline
    \end{tabularx}
  \end{adjustbox}
  \caption{The identified 26 papers with corresponding categories}
  \label{tab:mytable3}
\end{table}

\section{Results}\label{Results}

The categorization of 26 selected publications based on the codes (as discussed in the methodology section) enabled the author to classify the papers into 6 focus areas in which research is being conducted at ASSETS at the intersection of XR and disability (addressing \textbf{RQ1}). The 6 focus areas along with the papers addressing XR research in those areas is discussed in detail in Section \ref{Focus}. Moreover, the author identified that varied qualitative, quantitative, and mixed-methods were followed by the researchers and a detailed description of the same (addressing \textbf{RQ2}) is discussed in the Section \ref{Methods}. It also seemed important to perform high level statistics and analyze the temporal trends in research over the past 5 years in context to number of publications in assisitve XR research, XR technologies used, disabilities addressed at ASSETS, and methodologies used to identify the emerging trends and possible future directions in assistive XR research for disability as discussed in the Section \ref{Trends} (addressing \textbf{RQ3}). Supporting the review, the following sections, hence, address research questions \textbf{RQs 1,2 and 3} in detail.

\subsection{Focus areas of XR research at ASSETS (2019-2023)}\label{Focus}

\begin{figure}[htbp]
  \centering
  \includegraphics[width=\textwidth]{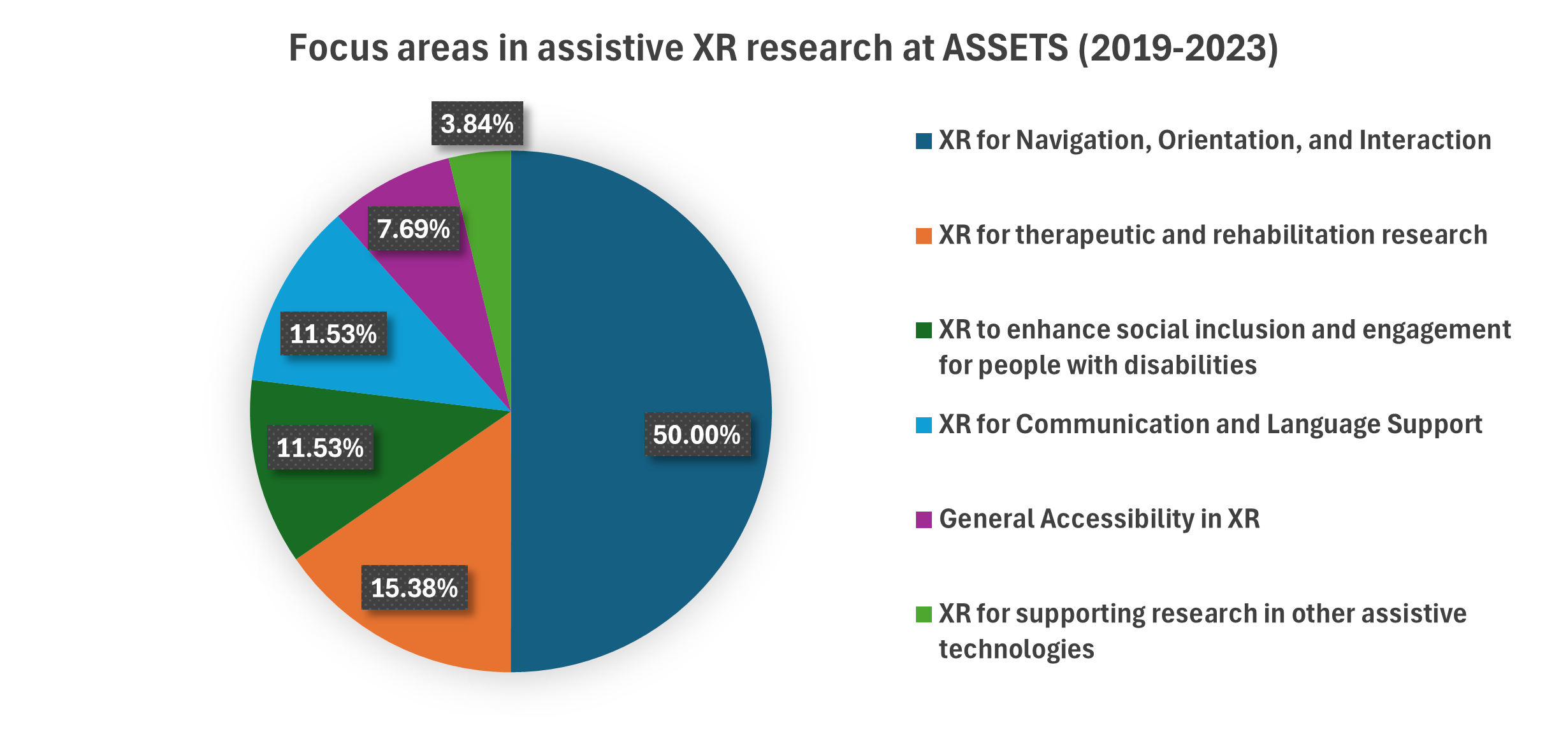}
  \caption{Key focus areas in assistive XR research explored at ASSETS from 2019-2023 (with percentage distribution)}
  \label{fig:focusareas_piechart}
  \Description{Pie graph showing percentage distribution of key focus areas in assistive XR research at ASSETS from 2019-2023.}
\end{figure}

Figure \ref{fig:focusareas_piechart} shows the 6 key focus areas explored in assistive XR research for disability in ACM ASSETS conference along with their percentages (as identified by the author).  Furthermore the following sections describe the focus areas in detail along with their statistics. 

\subsubsection{XR for Navigation, Orientation, and Interaction}

One of the major focal areas of research at ASSETS (2019-2023) (with 50\% of papers (N=13) focusing in this direction of research) was identified to be in the use of XR to enable people with disabilities "accessibly" navigate, orient, and interact in the physical world as well as the virtual environments \cite{anderton2022investigating} \cite{su2023demonstration}. For example, researchers in \cite{troncoso2020aiguide} \cite{yoon2019leveraging} created phone-based AR applications to help people with vision impairments to navigate and locate objects in indoor settings. Aldas et al.  \cite{troncoso2020aiguide} created, \textit{AI Guide}, a phone-based AR application that supports people with vision impairments to use their smartphones to locate and pick up objects around them. Yoon et al. built \textit{Clew} that allows users with vision impairments to record routes and load them later to get a haptic, speech, and sound feedback that guide them along the pre-recorded route. Pei et. al \cite{pei2023embodied} developed a VR technique, \textit{Embodied Exploration}, that facilitates people using wheelchairs to "virtually" tour and interact with unfamiliar places remotely using high-fidelity digital replicas of physical environments in VR through embodying avatars. Collins et al. \cite{collins2023guide} and Ji et al. \cite{ji2022vrbubble} explored ways to build inclusive social VR environments for people with vision impairments (Blind and Low-vision) through accessible techniques of navigation and orientation to interact with others in a social VR setting. \textit{VR Bubble} by Collins et. al \cite{collins2023guide} provides an audio-based stimuli to provide information about the identity, direction, and amount of avatars surrounding the user in a social VR setting. Ji et al. \cite{ji2022vrbubble} designed a framework where a physical guide can help people with vision impairments navigate in VR. Meanwhile, Franz et al. \cite{franz2023comparing} and Mahmud et al. \cite{mahmud2023eyes} conducted studies with people with limited mobility and upper-body impairments to identify accessible locomotion techniques in VR for people with sensorimotor disabilities. For instance,  Franz et al. \cite{franz2023comparing}
compared six locomotion techniques in terms of accessibility, workload, and experience factors, finding the "Teleport" technique to be the best technique to move and navigate in VR (while user experience and entertainment factors also played a key role). Mahmud et al. \cite{mahmud2023eyes} investigated visual feedback techniques to assist people with mobility impairments for maintaining balance in VR. Moreover, some researchers developed XR frameworks/applications encoded with apriori information for assisting designers in creating virtual environments, that can enable people with limited mobility orient themselves in VR. For instance, Franz et al. \cite{l2021nearmi} created \textit{Nearmi}, a framework that enhances accessible interaction techniques for people with sensorimotor disabilities, enabling them to \textit{gain awareness of POIs (point of interests) in virtual environments, and automatically re-orient the virtual camera toward a selected POI}. 
While the aforementioned works focused on sensorimotor disabilities and vision impairments, there has also been a focus on building spatial and interaction cues for Deaf and Hard of hearing (DOH) communities. Li et al. created \textit{SoundVizVR} application that generates visual representations of loudness, duration, and location of sound sources in VR, enabling the Deaf or Hard of hearing people to use these sound indicators to navigate and interact in VR.  

\subsubsection{XR for therapeutic and rehabilitation research}

The second area of focus with N=4 publications out of 26 (15.38\%) was identified to be use of XR for therapeutic and rehabilitation research. It was interesting to note that 50\% of the research in this focus area catered towards XR for people on the autism spectrum. Mcgowan and Mcgregor \cite{mcgowan2023investigation} conducted an online survey with over 200 autistic participants and caregivers, identifying that AR and VR technologies could be useful aids for reduction of unpredictable stress-related events. Meanwhile, Ahsen et al. \cite{ahsen2022designing} designed an AR application for therapists and educational professionals working in autistic contexts. The AR application built by the authors enabled the professionals to create AR experiences for conducting learning activities with autistic children. Chao and Peiris \cite{chao2022college} explored virtual reality therapy (or VRT) to be adopted at college campuses opening opportunities for remote counselling services. Furthermore, Riaz et al. \cite{riaz2020exploratory} did an exploratory study to identify the challenges faced by older adults (diagnosed with Aphasia) and provide effective support using VR by aiding in language and speech production.

\subsubsection{XR to enhance social inclusion and engagement for people with disabilities}

Another major focus of research (N=3 out of 26; 11.53\%) as identified by the author was the use of XR for enhancing inclusion of people with disabilities among social groups in the virtual space. Gualano et al. \cite{gualano2023invisible} explored how people with invisible disabilities (e.g. chronic health conditions) want to represent and disclose their disabilities on social VR platforms. The authors suggested avatar customization as per preferences of people with invisible disabilities. Similarly, Zhang et al. \cite{zhang2022s} conducted a systematic review of fifteen social VR applications to investigate how people with disabilities prefer representation of their avatars in VR. Furthermore, Zhang et al. \cite{zhang2023diary} conducted a diary study with people with disabilities (PWDs) to study how disability disclosure via the avatars would affect PWDs social and interaction dynamics with others in social VR application such as VRChat. Interestingly, no study was conducted to study social co-presence of PWDs in an AR setting. 

\subsubsection{XR for Communication and Language Support}

Another area of focus at ASSETS in XR research for disability was identified to be supporting Deaf and Hard of hearing (DOH) communities for language and communication support using XR technologies (N=3 out of 26; 11.53\%). For instance, Quandt et al. \cite{quandt2020teaching} developed a proof-of-concept system where avatars (animated virtual humans recorded using motion-capture data) would provide introductory ASL (American Sign Language) sessions in VR. Mathhew et al. \cite{mathew2022access}, on the other hand, explored how \textit{Access on Demand (AoD)}, an AR application (on AR smart glasses) could be beneficial for Deaf or Hard of Hearing communities by providing them on-demand real-time captioning and sign language interpretation services. Meanwhile, researchers such as Berke et al. \cite{berke2020chat} explored hands-free communication support by developing a custom developed prototype (a wearable hat) where DOH could do remote video interpretation for communicating with hearing peers without using a hand-held phone, enabling them to do multitasking and indulge in a natural conversation. 

\subsubsection{General Accessibility in XR}

While most of the papers published at ASSETS in the time range of 2019-2023 focused and addressed a particular XR functionality or use-case for accessibility in XR, a few (N=2 out of 26, 7.69\%) papers addressed the overall accessibility in such XR systems. Mott et al. \cite{mott2020just} conducted semi-structured interviews with people with mobility limitations to understand the accessibility issues they face in VR. The authors identified a list of seven barriers related to physical accessibility that people with mobility limitations might encounter while using VR devices. Herskovitz et al. \cite{herskovitz2020making} analyzed 105 existing mobile AR applications (for iOS) and identified challenges (e.g. locating and placing objects in AR) and areas of future research for making AR fully accessible (e.g. continous involvement of target communities in design decisions). 

\subsubsection{XR for supporting research in other assisitve technologies}

The last focus group (N=1 out of 26, 3.84\%) identified was the use of XR to identify challenges in the development of other assistive technologies. Palaniappan et al. \cite{palaniappan2019identifying} used VR technology to identify comfort areas in 3D space for persons with upper extremity mobility impairments. The authors created a VR exertion game to identify areas in space where people with upper body sensorimotor disabilities would be comfortable in doing frequent motions, hence, spotting the areas where other assistive technologies could be placed ergonomically. 

\subsection{Research methods adopted for assistive XR research at ASSETS}\label{Methods}

Given the human-centred design approach followed by most researchers at ASSETS, various research methods were adopted for conducting XR research for disability. Researchers employed qualitative (69.23\%), quantitative (3.84\%) as well as mixed-methods approach (26.92\%) to evaluate their results (please refer to Figure \ref{methodologies_piechart}). The sections below give a detailed overview of the approaches  followed in these three categories.

\subsubsection{Qualitative methods for XR research at ASSETS} Overall 69.23\% researchers employed qualitative methods to evaluate their research on assistive XR technologies for people with disabilities. However, the most common qualitative method employed by HCI researchers was conducting semi-structured interviews (66.67\% of researchers who used qualitative methods conducted semi-structured interviews). As a part of user studies, researchers conducted semi-structured interviews with participants to: evaluate their prototypes \cite{su2023demonstration} \cite{pei2023embodied} \cite{anderton2022investigating} \cite{troncoso2020aiguide}, to understand the experiences of users (i.e. people with disabilities) in XR \cite{gualano2023invisible}  \cite{franz2023comparing} \cite{zhang2022s} \cite{riaz2020exploratory} \cite{mott2020just}, and investigate use of XR for therapy and rehabilitation \cite{mcgowan2023investigation} \cite{ahsen2022designing} \cite{chao2022college}. Researchers also performed user interviews followed by thematic analysis to identify major themes/categories in the responses of participant post interview sessions \cite{zhang2023diary} \cite{pei2023embodied} \cite{ahsen2022designing} \cite{zhang2022s} \cite{l2021nearmi} \cite{herskovitz2020making}. For example, Gualano et al. \cite{gualano2023invisible} used thematic analysis in their paper to explore how people with disabilities (PWDs) want to represent themselves in Social VR situations. The researchers identified that stigmas around disabilities,  presence of disabled groups in virtual spaces, and possibilities to customize their avatar representations can widely influence PWDs decisions on self-representing themselves in social virtual spaces. 
Researchers also used various scales such as Likert-Scale (5-point and 7-point) and System Usability Scale (SUS) for subjective analysis of their questionnaires to support their studies. While some groups such as \cite{collins2023guide} \cite{li2022soundvizvr} \cite{ji2022vrbubble} \cite{berke2020chat} adopted the 5-point scale, a few \cite{pei2023embodied} \cite{anderton2022investigating}  \cite{herskovitz2020making} used 7-point scale to evaluate their studies. NASA-TLX work load questionnaire \cite{hart1988development} was also used by some researchers \cite{franz2023comparing} \cite{li2022soundvizvr} \cite{h2020clevr} to analyze the subjective cognitive-load measures as participants used their VR/AR applications. Finally, researchers like Zhang et al. \cite{zhang2023diary} and Ahsen \cite{ahsen2022designing} also used diary-studies to evaluate their explorations in VR and AR respectively for disabled communities. 

\subsubsection{Quantitative methods for XR research at ASSETS}

While several quantitative methods were adopted for mixed-method studies, that is, integration of quantitative measures with qualitative studies, Palaniappan et al. \cite{palaniappan2019identifying} focused only on quantitative analysis such as the use of machine learning techniques (e.g. clustering algorithms) to evaluate comfort areas in 3D space for persons with upper extremity mobility impairments using virtual reality. Hence, adoption of just quantitative measures for XR research for people with disabilities was found to be rare (3.38\%) at ACM ASSETS for the years 2019-2023.

\subsubsection{Mixed-methods for XR research at ASSETS}

Another prevalent research method adopted by HCI researchers at ASSETS (2019-2023) was mixed-methods with N=7 papers (26.92\%) employing the same \cite{berke2020chat} \cite{yoon2019leveraging} \cite{li2022soundvizvr} \cite{franz2023comparing}. For instance, Mahmud et al. \cite{mahmud2023eyes} in their investigation of visual feedback (in VR) to improve gait for people with mobility impairments (MI), firstly, conducted qualitative questionnaires (ABC forms \cite{powell1995activities} and SSQ \cite{kennedy1993simulator}) and later performed quantitative studies and analysis on collected data such as various gait parameters (e.g. walking velocity, step length etc.). Such statistical analysis was conducted with both people with and without MI using a mixed-model ANNOVA and t-tests to identify the significant differences between the two groups. Similarly, Aldas et al. \cite{troncoso2020aiguide} for creating their AR application, \textit{AIGuide}, firstly, did a quantitative analysis on the video chat record data collected from the participant trials. Quantitative evaluation included performance and effectiveness measures by calculation of time each participant took to complete the trials (using their application). Later the researchers conducted a semi-structured interview to qualitatively evaluate the effectiveness of their interface.  

\subsection{Statistics and temporal trends in assistive XR research at ASSETS (2019-2023)}\label{Trends}

\subsubsection{Number of research articles submitted at ASSETS}

\begin{figure}[htbp]
  \centering
  \includegraphics[width=0.7\textwidth]{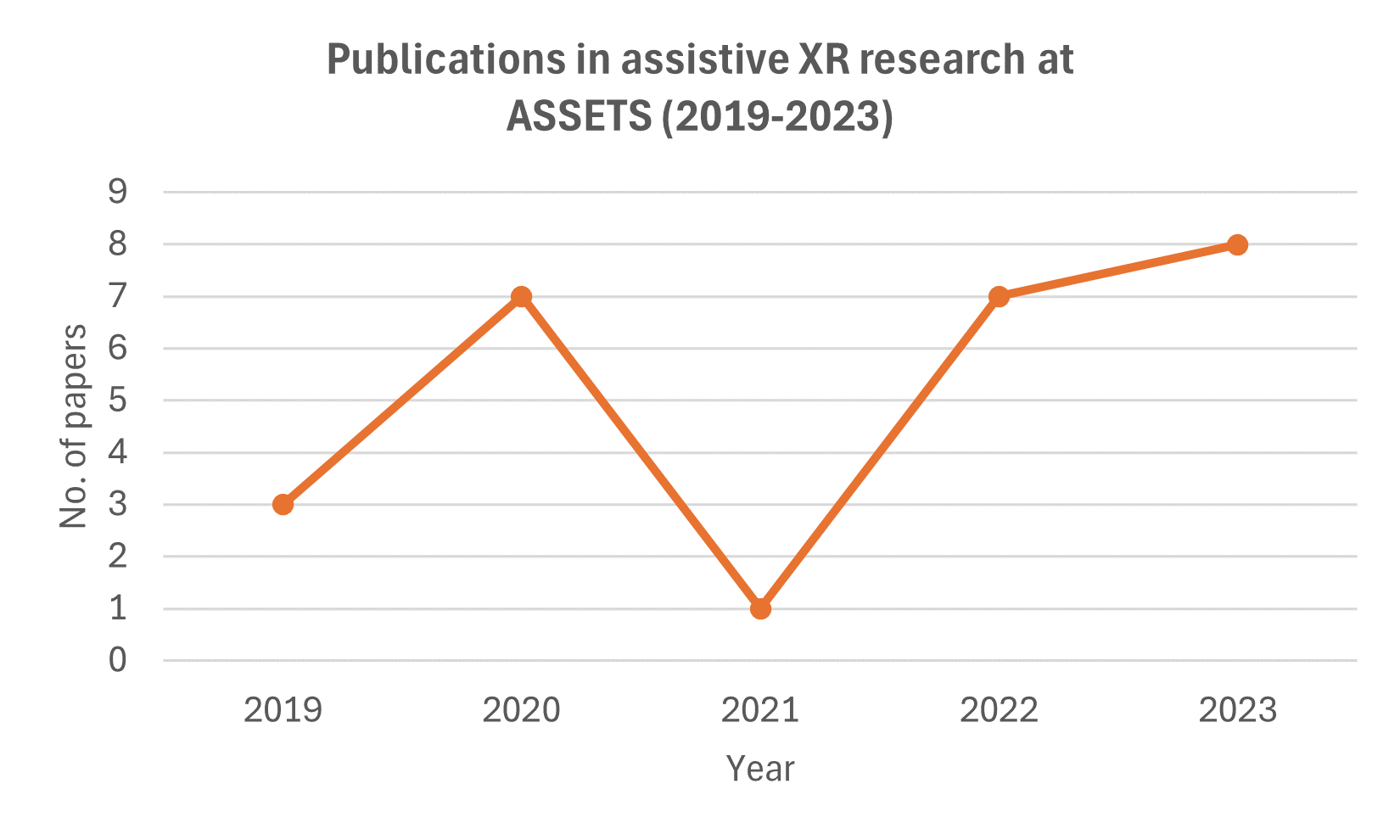}
  \caption{Publications in assistive XR research at ASSETS (2019-2023)}
  \label{publications_temporaltrend}
  \Description{Line graph showing temporal trend in the number of publications in assistive XR research at ASSETS from 2019 to 2023.}
\end{figure}

Screening articles from 2019-2023 published at ACM ASSETS, the author identified an evolving interest in assistive XR research for disabled communities. Figure \ref{publications_temporaltrend} shows the number of papers published at ACM ASSETS over the period of 2019-2023. The screening process reflected that the number of papers using Virtual and Augmented Reality technologies in context of disability almost tripled from 2019 (3 publications) to 2023 (8 publications). It is also noticeable that there is a sudden fall in assistive XR research at ASSETS 2021. The author suspects the reason being the halting of user studies during the COVID-19 pandemic (academic year 2020-2021), as stated by other researchers working in the field of HCI \cite{kroma2022reality}. Estrada and Prasolova-Forland running an XR lab at a Norwegian university share in \cite{garcia2022running} that "\textit{execution of studies requiring volunteers became challenging due to the restrictions on physical contact and meeting}" during the pandemic. However, since 2022, the research is ever growing with almost equivalent numbers of papers published (7 in 2022 to 8 in 2023). With the launch of more consumer-targeted headsets such as Quest 3 in 2023 and Apple Vision Pro in 2024, the author predicts increase in XR research at ASSETS for and with people with disabilities.

\subsubsection{Use of XR technologies at ASSETS}

\begin{figure}[htbp]
  \centering
  \begin{subfigure}[b]{0.45\textwidth}
    \centering
    \includegraphics[width=\textwidth]{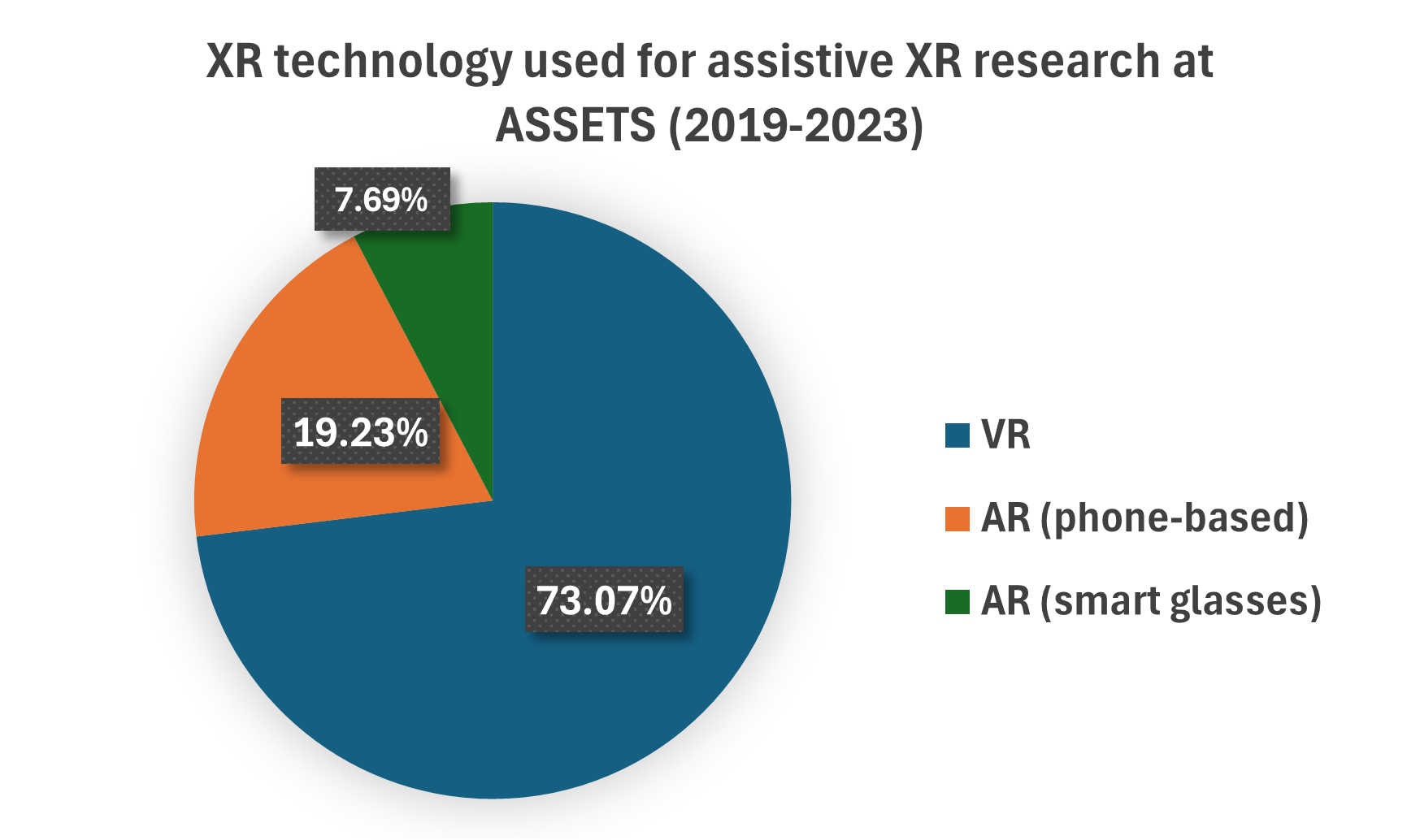}
    \caption{Percentage distribution of XR technologies used for assistive XR research at ASSETS}
    \label{xrtechnologyused_piechart}
    \Description{Pie graph showing percentage distribution of XR technologies (VR, AR (phone-based), AR (smart glasses)) used in assistive XR research at ASSETS from 2019-2023.}
  \end{subfigure}
  \hfill
  \begin{subfigure}[b]{0.5\textwidth}
    \centering
    \includegraphics[width=\textwidth]{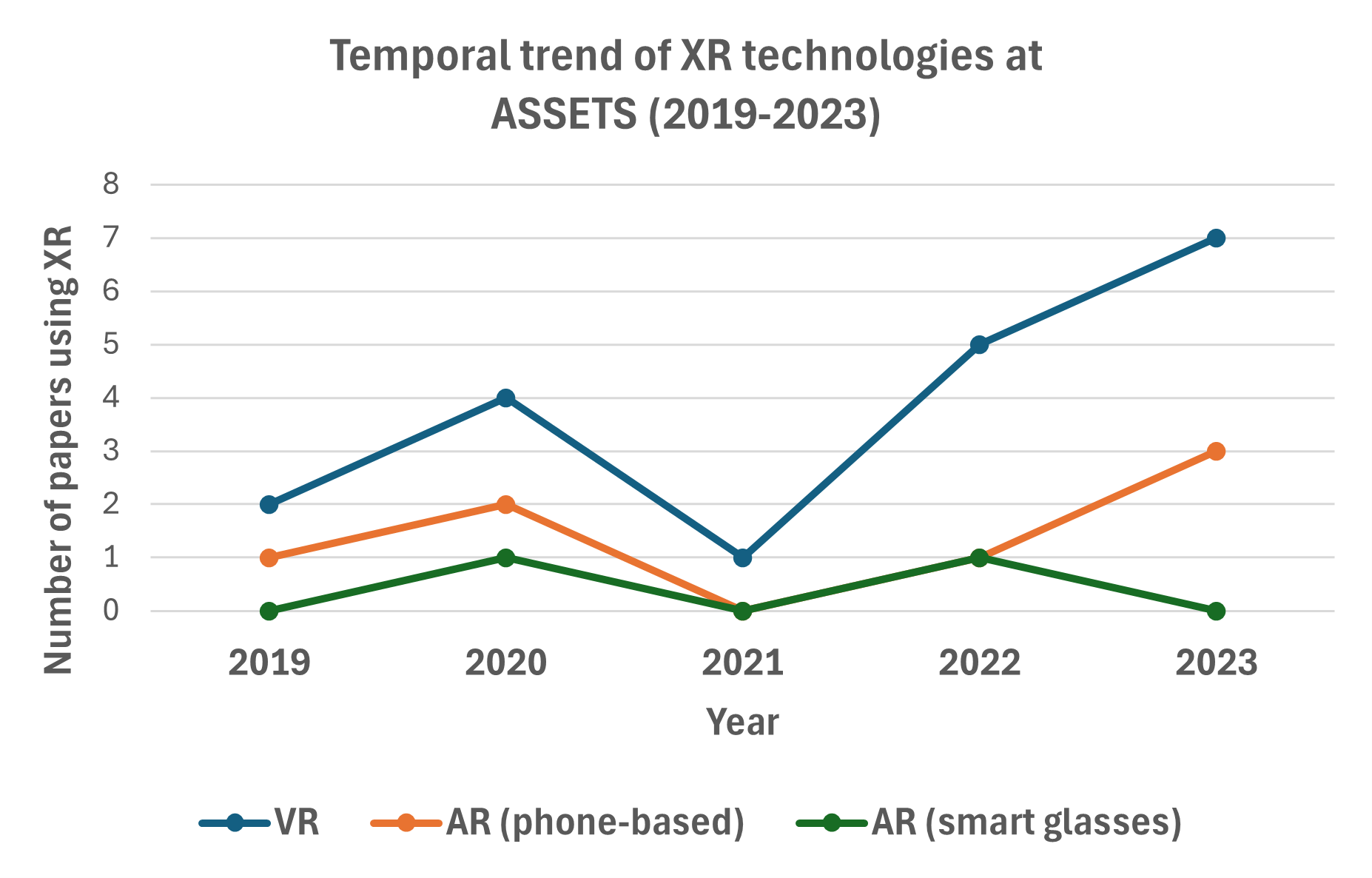}
    \caption{Temporal trend in the use of XR Technology at ASSETS}
    \label{xrtechnologyused_temporaltrend}
    \Description{Line graph showing temporal trend in XR technology usage at ASSETS from 2019-2023, with VR, AR (phone-based), and AR (smart glasses) depicted by colored lines}
  \end{subfigure}
  \caption{Patterns and temporal trends in use of XR technologies at ASSETS}
  \label{trends_patterns_xrtechnologyused}
  \Description{A pie graph and line graph depicting the statistics and temporal trends of XR technologies used in assistive XR research at ASSETS from 2019 to 2023.}
\end{figure}

The researchers working in the field of assistive XR research used three types of XR technologies: Head-mounted displays (for VR research), smart-glasses, and smart-phones (for AR research).
The majority of research (almost 73\%) was conducted in VR applications and solutions for people with disabilities (with N=19 papers on VR). Around 27\% of researchers focused on AR-based solutions for people with disabilities. However, in context of AR oriented solutions, researchers mostly used phone-based AR applications (with 19.23\% of overall publications in between 2019-2023), while two papers used Vuzix Blade smart glasses for AR applications (with 7.69\% of research). Figure \ref{xrtechnologyused_piechart} shows the percentage distribution of the XR technologies explored at ASSETS from 2019-2023. 

Reviewing the temporal trend in the use of XR technologies at ASSETS, it was noticed that the interest in investigation of accessible "VR solutions" for people with disabilities at ASSETS has been tripled over the last 5 years (from 2 papers in 2019 to 7 papers in 2023) as depicted in Figure \ref{xrtechnologyused_temporaltrend}. As the interest in AR is increasing at ASSETS for assisitve XR research, the use of phone-based AR shows an increasing trend while the use of smart glasses for AR stays constant with none of the researchers using such AR technologies in the year 2019, 2021, and 2023. With the rise of XR technologies such as Quest 2 and 3 offering passthrough functionality (that enables the use of same head-mounted display for both VR and AR), it is predictable that assisitve XR research might show a hike in AR research at the forthcoming ASSETS conferences.

\subsubsection{Disabilities addressed at ASSETS}

\begin{figure}[htbp]
  \centering
  \includegraphics[width=0.9\textwidth]{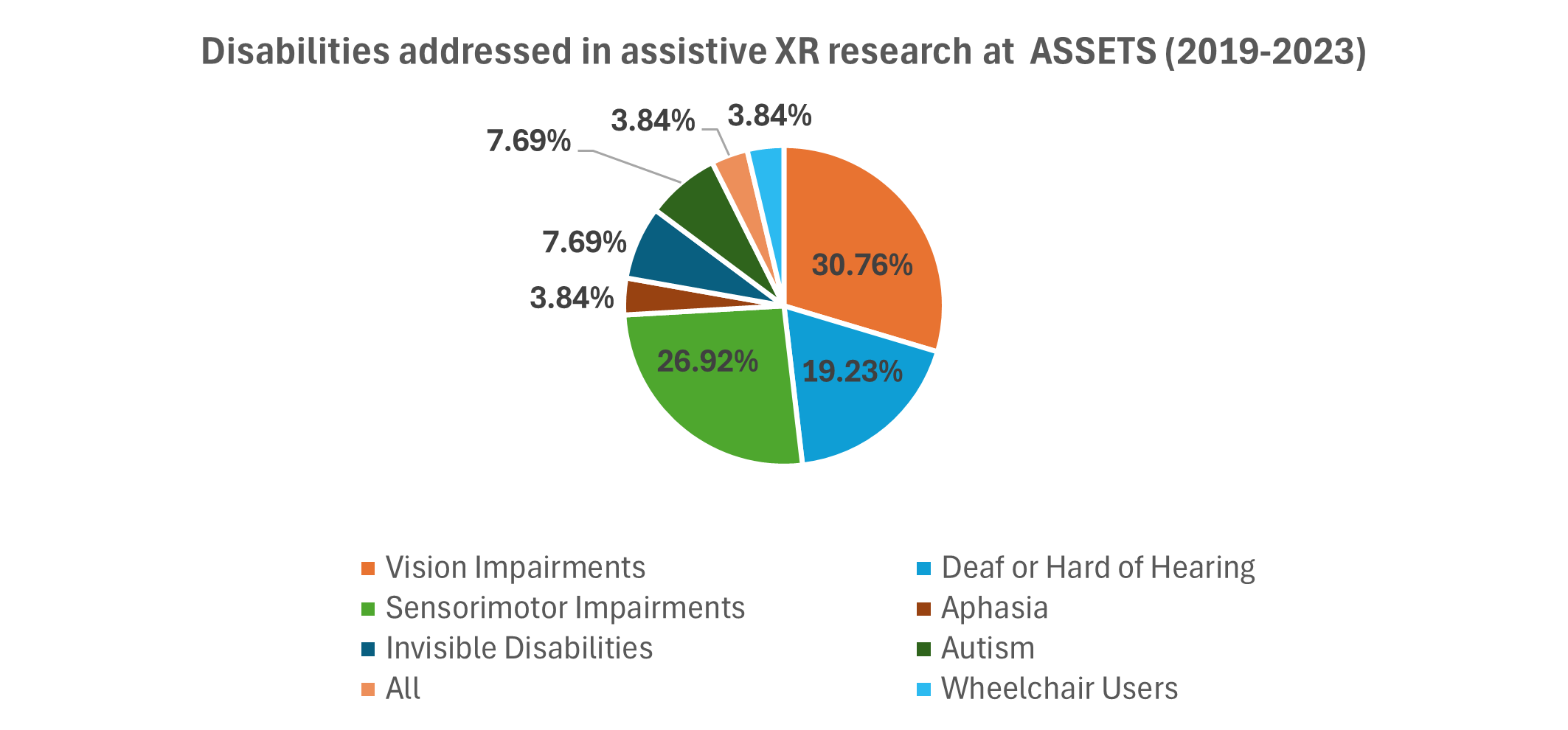}
  \caption{Percentage distribution of disabilities addressed in assistive XR research at ASSETS (2019-2023)}
  \label{disabilities_piechart}
  \Description{Pie graph displaying percentage distribution of disabilities addressed in assistive XR research at ASSETS (2019-2023)}
\end{figure}

A wide range of disabilities were addressed in context of building accessible XR solutions for disabled communities (as shown in Figure \ref{disabilities_piechart}). The most common disabilities addressed by the researchers were people with vision impairments, groups with limited mobility (or sensorimotor impairments), and people who are deaf or hard of hearing. As shown in Figure \ref{disabilities_piechart}, 30.76\% of papers targeted people with visual disabilities (low vision, blindness etc.) and 26.92\% explored XR technologies for people with limited mobility such as upper body sensorimotor impairments. 19.23\% of researchers explored XR for deaf or hard of hearing communities. Other disabilities which were addressed but in low numbers were Aphasia (3.84\%) and users of wheelchairs (e.g. older adults) (3.84\%). XR technologies for invisible disabilities such as neuro-divergent communities were also explored with 7.69\% of papers published in ASSETS (2019-2023) targeted such groups. Many researchers also explored the use of XR for therapy and rehabilitation research for people with autism (7.69\%). A few papers also explored the general accessibility concerns in XR technologies targeting all disabled groups (3.84\% papers targeted all disabilities).

\begin{figure}[htbp]
  \centering
  \includegraphics[width=0.85\textwidth]{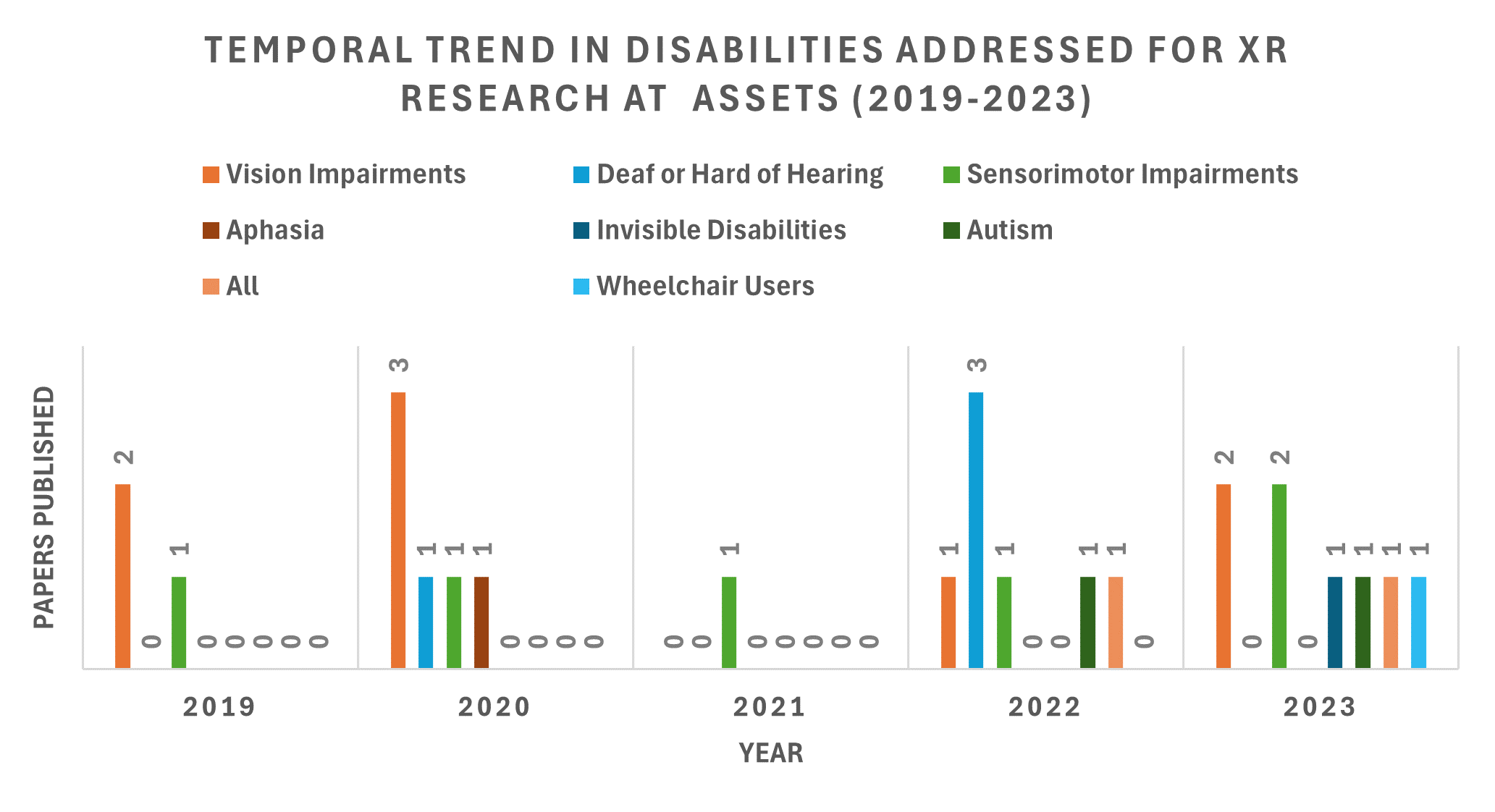}
  \caption{Temporal trend in disabilities addressed in assistive XR research at ASSETS (2019-2023)}
  \label{disabilitiesaddressed_temporaltrend}
  \Description{Collection of 5 bar graphs titled "Temporal trend in disabilities addressed in assistive XR research at ASSETS (2019-2023)", each representing disability categories from 2019 to 2023.}
\end{figure}

Furthermore, Figure \ref{disabilitiesaddressed_temporaltrend} shows the variation in the disabilities addressed for assisitve XR research in the timeline of 2019-2023. Research in XR for people with vision impairments (e.g blindness or low-vision) remains constant with 1-3 papers addressing visual disabilities every year from 2019 to 2023 (except 2021, which is suspected due a decline in papers due to COVID-19 pandemic). ACM ASSETS witnessed an increasing research in XR for deaf and hard of hearing (DOH) communities with the highest number of publications (3 papers) in 2022 addressing DOH communities. It is also interesting to note that every year (from 2019-2023), atleast 1 paper was addressed to people with limited mobilities (or sensorimotor impairments). The interest of researchers in exploring XR for people with sensorimotor impairments has increased, as reflected by the hike in the numbers of papers from 1 in 2019 to 2 in 2023. It was also inferred that research was also conducted to explore accessible XR for people using wheelchairs (e.g. older adults). While autistic communities were not addressed in assisitve XR research from 2019-2021, the last two years (2022 and 2023) witness a contant number of publications addressing autistic communities (1 paper both in 2022 and 2023).

\begin{figure}[htbp]
  \centering
  \begin{subfigure}[b]{0.5\textwidth}
    \centering
    \includegraphics[width=\textwidth]{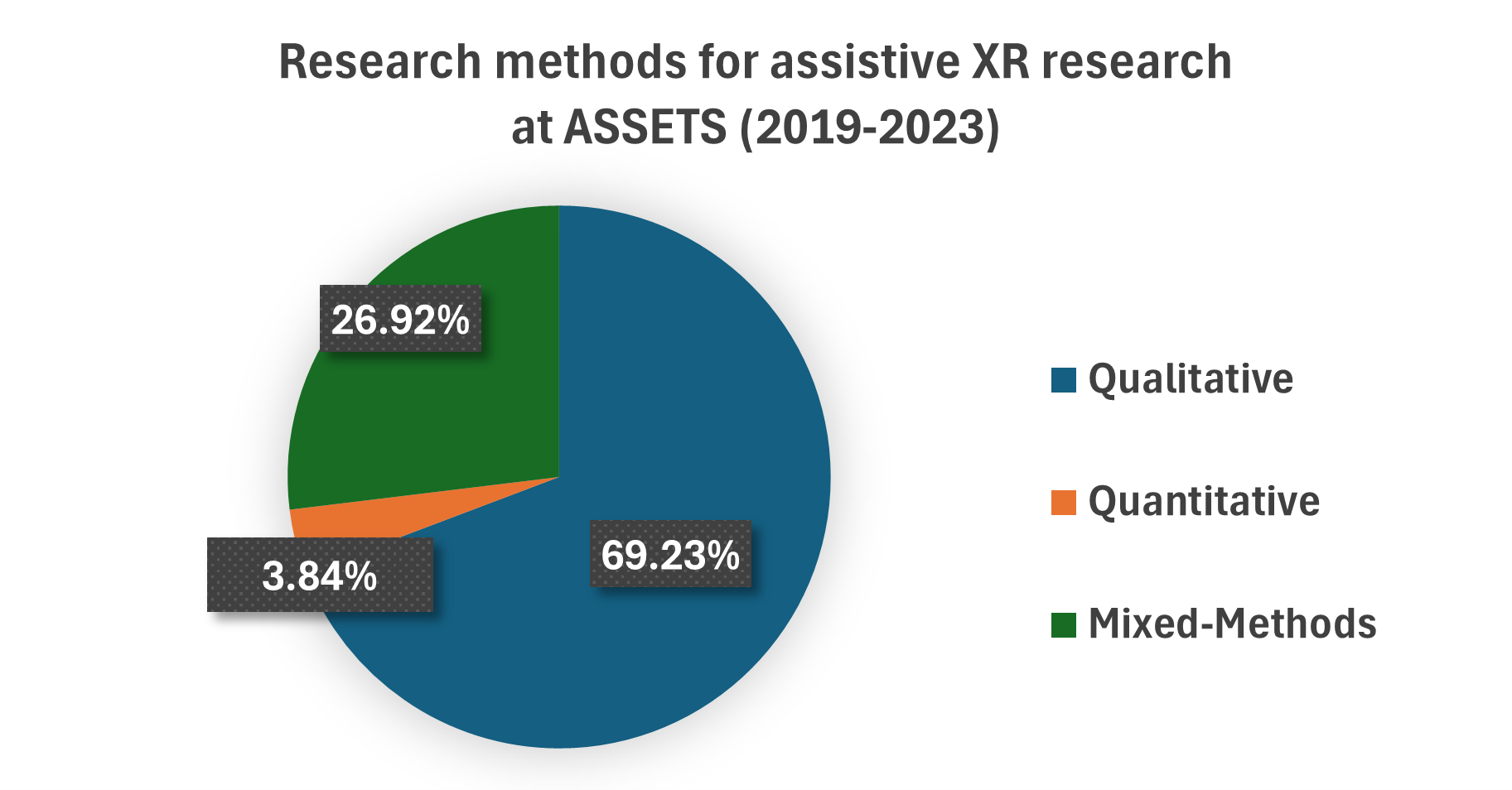}
    \caption{Percentage distribution of methodologies used in assistive XR research at ASSETS (2019-2023)}
    \label{methodologies_piechart}
    \Description{A pie graph depicting the percentage distribution of methodologies (Qualitative, Quantitative, and Mixed-Methods) used in assistive XR research at ASSETS (2019-2023).}
  \end{subfigure}
  \hfill
  \begin{subfigure}[b]{0.5\textwidth}
    \centering
    \includegraphics[width=\textwidth]{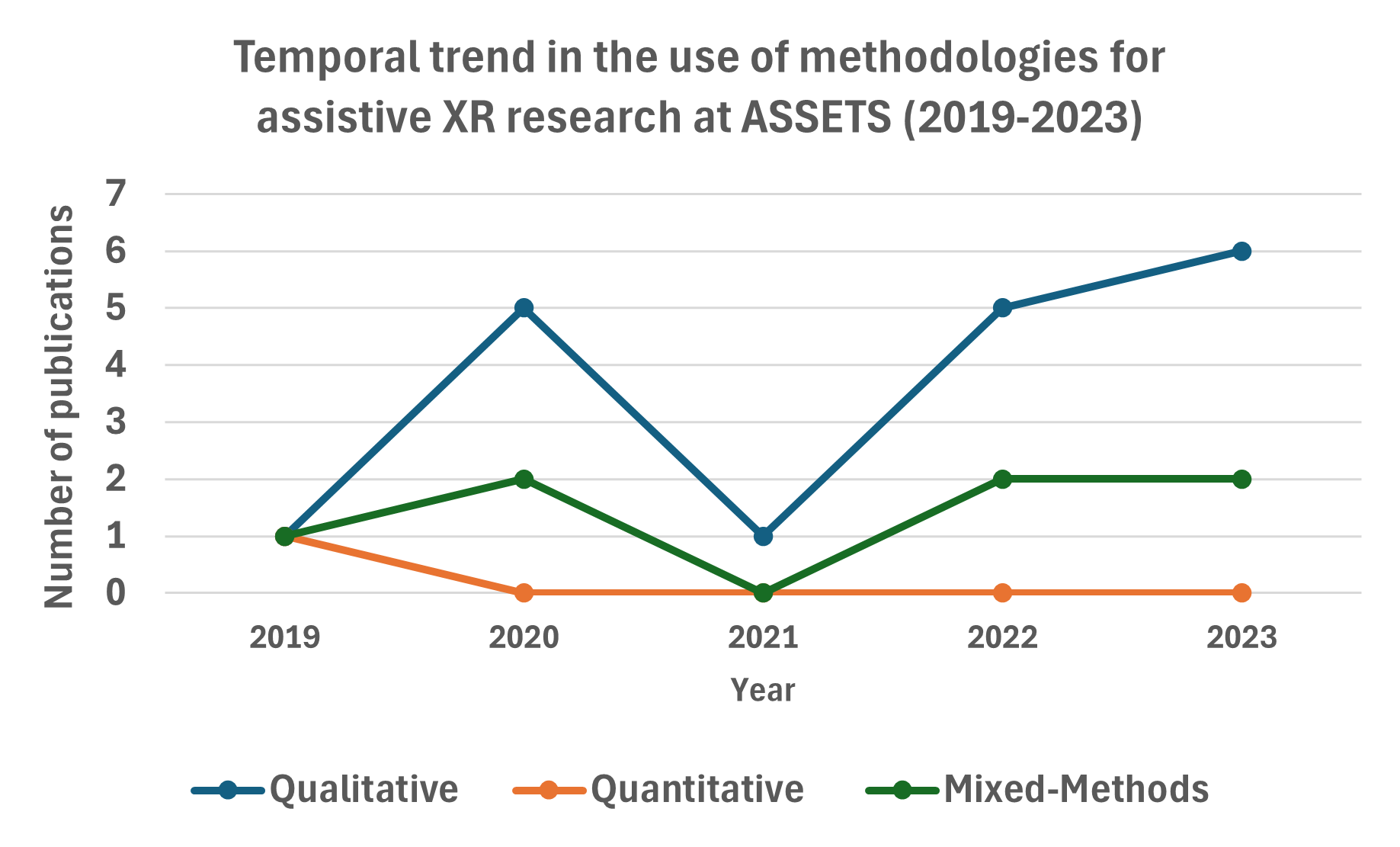}
    \caption{Temporal trend in the use of methodologies for assistive XR research at ASSETS (2019-2023)}
    \label{methodologies_temporaltrend}
    \Description{A line graph showing three lines indicating the usage of Qualitative, Quantitative, and Mixed-Methods in assistive XR research at ASSETS over the years 2019-2023.}
  \end{subfigure}
  \caption{Patterns and temporal trends in use of XR technologies at ASSETS (2019-2023)}
  \label{trends_patterns_methodologies}
  \Description{A pie graph and line graph depicting the statistics and temporal trends of XR methodologies used in assistive XR research at ASSETS from 2019 to 2023.}
\end{figure}

\subsubsection{Methodologies for assistive XR research at ASSETS}

There were three research methodologies employed by researchers for evaluating their studies for assisitve XR research at ASSETS in the time range of 2019-2023 (as shown in Figure \ref{methodologies_piechart}). The most common research methods used included qualitative analysis (with 69.23\% of papers using qualitative methods such as semi-structured interviews etc.). Use of just quantitative assessment to evaluate XR applications for people with disabilities was found to be rare (with 3.84\% publications). Around 26.92\% research employed mixed-methods for assisitve XR reach at ASSETS. Figure \ref{methodologies_temporaltrend} shows the temporal variation is the use of aforementioned methodologies over the time period 2019-2023 at ACM ASSETS.

The use of qualitative methods hiked in the last 5 years, from 1 group of researchers using the method in 2019 to highest count of papers published in 2023 (N=6) choosing qualitative methods for evaluations. Researchers stopped using quantitative methods as the only means of evaluation in the last few years (with 1 publication using it in 2019 to none till 2023). However, quantitative evaluation was integrated with qualitative assessment in the form of mixed-method studies over the time range of 2019-2023 at ASSETS, with at-least 1-2 publications using mixed-method studies every year in the past 5 years. With the human-centred approach of researchers at ASSETS to create accessible solutions for disabled communities, the use of qualitative studies (which capture the subjective view points of users) for evaluation (or integrated with quantitative analysis) thus seems evident at upcoming ASSETS conferences.

\section{Conclusion}

This work presents a scoping review of previous research conducted at ACM ASSETS conference (2019-2023) in the context of assistive XR research for disability. Through a pool of 1595 articles, 26 articles are identified that specifically focus on XR research for people with disabilities. A detailed analysis on the 26 articles revealed 6 key focus areas that are currently being explored at ASSETS in assistive XR research for disability. The paper describes the key focus areas and their statistics in detail. Furthermore, the review reports three methodologies (qualitative, quantitative, and mixed-methods) adopted by researchers and provides a detailed statistical and descriptive overview of all the methods used under each category in detail. Lastly, the review captures the statistics and temporal trends regarding the number of publications, XR technologies used, disabilities addressed, and methodologies adopted for assistive XR research at ASSETS. The trends revealed an increase in the number of papers published in assistive XR research with most studies focusing on VR technologies. Moreover, it was deduced that the majority of studies were targeted for people with vision impairments with qualitative methods as the main research methodology used for conducting such studies.  

\section{Limitations and Future work}

The author acknowledges the limitation of this work. Since, this work was conducted by a single author, a two phase-screening could not be carried out due to high number of research articles selected for the screening process (with 520 papers screened). This could have led to some biases in data collection and hence overall analysis. The author is also aware that there are other ACM Conferences that also address research at the intersection of HCI, XR, and disability (e.g CHI, UbiComp etc.), inclusion of which could have enriched this scoping review to address overall assisitve XR research themes prevalent in the HCI community. However, the goal of this paper was to provide a detailed overview of assistive XR research at ASSETS. The author does aim to intgrate these results with previous work conducted in other ACM conferences as a future work. 

\bibliographystyle{ACM-Reference-Format}
\bibliography{sample-base}

%%% -*-BibTeX-*-
%%% Do NOT edit. File created by BibTeX with style
%%% ACM-Reference-Format-Journals [18-Jan-2012].

\begin{thebibliography}{59}

%%% ====================================================================
%%% NOTE TO THE USER: you can override these defaults by providing
%%% customized versions of any of these macros before the \bibliography
%%% command.  Each of them MUST provide its own final punctuation,
%%% except for \shownote{}, \showDOI{}, and \showURL{}.  The latter two
%%% do not use final punctuation, in order to avoid confusing it with
%%% the Web address.
%%%
%%% To suppress output of a particular field, define its macro to expand
%%% to an empty string, or better, \unskip, like this:
%%%
%%% \newcommand{\showDOI}[1]{\unskip}   % LaTeX syntax
%%%
%%% \def \showDOI #1{\unskip}           % plain TeX syntax
%%%
%%% ====================================================================

\ifx \showCODEN    \undefined \def \showCODEN     #1{\unskip}     \fi
\ifx \showDOI      \undefined \def \showDOI       #1{#1}\fi
\ifx \showISBNx    \undefined \def \showISBNx     #1{\unskip}     \fi
\ifx \showISBNxiii \undefined \def \showISBNxiii  #1{\unskip}     \fi
\ifx \showISSN     \undefined \def \showISSN      #1{\unskip}     \fi
\ifx \showLCCN     \undefined \def \showLCCN      #1{\unskip}     \fi
\ifx \shownote     \undefined \def \shownote      #1{#1}          \fi
\ifx \showarticletitle \undefined \def \showarticletitle #1{#1}   \fi
\ifx \showURL      \undefined \def \showURL       {\relax}        \fi
% The following commands are used for tagged output and should be
% invisible to TeX
\providecommand\bibfield[2]{#2}
\providecommand\bibinfo[2]{#2}
\providecommand\natexlab[1]{#1}
\providecommand\showeprint[2][]{arXiv:#2}

\bibitem[see({[n.\,d.]})]%
        {seevividly}
 \bibinfo{year}{[n.\,d.]}\natexlab{}.
\newblock \bibinfo{title}{Seevivdly}.
\newblock \bibinfo{howpublished}{\url{https://www.seevividly.com/}}.
\newblock
\newblock
\shownote{Accessed: 2024-04-23}.


\bibitem[xra({[n.\,d.]})]%
        {xraccess}
 \bibinfo{year}{[n.\,d.]}\natexlab{}.
\newblock \bibinfo{title}{XR Access Symposium}.
\newblock \bibinfo{howpublished}{\url{https://xraccess.org/symposium/}}.
\newblock
\newblock
\shownote{Accessed: 2024-04-23}.


\bibitem[Ahsen et~al\mbox{.}(2022)]%
        {ahsen2022designing}
\bibfield{author}{\bibinfo{person}{Tooba Ahsen}, \bibinfo{person}{Christina Yu}, \bibinfo{person}{Amanda O'Brien}, \bibinfo{person}{Ralf~W Schlosser}, \bibinfo{person}{Howard~C Shane}, \bibinfo{person}{Dylan Oesch-Emmel}, \bibinfo{person}{Eileen~T Crehan}, {and} \bibinfo{person}{Fahad Dogar}.} \bibinfo{year}{2022}\natexlab{}.
\newblock \showarticletitle{Designing a Customizable Picture-Based Augmented Reality Application For Therapists and Educational Professionals Working in Autistic Contexts}. In \bibinfo{booktitle}{\emph{Proceedings of the 24th International ACM SIGACCESS Conference on Computers and Accessibility}}. \bibinfo{pages}{1--16}.
\newblock


\bibitem[Anderton(2022)]%
        {anderton2022investigating}
\bibfield{author}{\bibinfo{person}{Craig Anderton}.} \bibinfo{year}{2022}\natexlab{}.
\newblock \showarticletitle{Investigating Sign Language Interpreter Rendering and Guiding Methods in Virtual Reality 360-Degree Content}. In \bibinfo{booktitle}{\emph{Proceedings of the 24th International ACM SIGACCESS Conference on Computers and Accessibility}}. \bibinfo{pages}{1--6}.
\newblock


\bibitem[Arksey and O'malley(2005)]%
        {arksey2005scoping}
\bibfield{author}{\bibinfo{person}{Hilary Arksey} {and} \bibinfo{person}{Lisa O'malley}.} \bibinfo{year}{2005}\natexlab{}.
\newblock \showarticletitle{Scoping studies: towards a methodological framework}.
\newblock \bibinfo{journal}{\emph{International journal of social research methodology}} \bibinfo{volume}{8}, \bibinfo{number}{1} (\bibinfo{year}{2005}), \bibinfo{pages}{19--32}.
\newblock


\bibitem[Berke et~al\mbox{.}(2020)]%
        {berke2020chat}
\bibfield{author}{\bibinfo{person}{Larwan Berke}, \bibinfo{person}{William Thies}, {and} \bibinfo{person}{Danielle Bragg}.} \bibinfo{year}{2020}\natexlab{}.
\newblock \showarticletitle{Chat in the hat: A portable interpreter for sign language users}. In \bibinfo{booktitle}{\emph{Proceedings of the 22nd International ACM SIGACCESS Conference on Computers and Accessibility}}. \bibinfo{pages}{1--11}.
\newblock


\bibitem[Biswas et~al\mbox{.}(2021)]%
        {biswas2021adaptive}
\bibfield{author}{\bibinfo{person}{Pradipta Biswas}, \bibinfo{person}{Pilar Orero}, \bibinfo{person}{Manohar Swaminathan}, \bibinfo{person}{Kavita Krishnaswamy}, {and} \bibinfo{person}{Peter Robinson}.} \bibinfo{year}{2021}\natexlab{}.
\newblock \showarticletitle{Adaptive accessible AR/VR systems}. In \bibinfo{booktitle}{\emph{Extended Abstracts of the 2021 CHI Conference on Human Factors in Computing Systems}}. \bibinfo{pages}{1--7}.
\newblock


\bibitem[Chao and Peiris(2022)]%
        {chao2022college}
\bibfield{author}{\bibinfo{person}{Vanny Chao} {and} \bibinfo{person}{Roshan Peiris}.} \bibinfo{year}{2022}\natexlab{}.
\newblock \showarticletitle{College students’ and campus counselors’ attitudes toward teletherapy and adopting virtual reality (Preliminary exploration) for campus counseling services}. In \bibinfo{booktitle}{\emph{Proceedings of the 24th International ACM SIGACCESS Conference on Computers and Accessibility}}. \bibinfo{pages}{1--4}.
\newblock


\bibitem[Chowdhury et~al\mbox{.}(2019)]%
        {chowdhury2019vr}
\bibfield{author}{\bibinfo{person}{Tanvir~Irfan Chowdhury}, \bibinfo{person}{Sharif Mohammad~Shahnewaz Ferdous}, {and} \bibinfo{person}{John Quarles}.} \bibinfo{year}{2019}\natexlab{}.
\newblock \showarticletitle{VR disability simulation reduces implicit bias towards persons with disabilities}.
\newblock \bibinfo{journal}{\emph{IEEE transactions on visualization and computer graphics}} \bibinfo{volume}{27}, \bibinfo{number}{6} (\bibinfo{year}{2019}), \bibinfo{pages}{3079--3090}.
\newblock


\bibitem[Collins et~al\mbox{.}(2023)]%
        {collins2023guide}
\bibfield{author}{\bibinfo{person}{Jazmin Collins}, \bibinfo{person}{Crescentia Jung}, \bibinfo{person}{Yeonju Jang}, \bibinfo{person}{Danielle Montour}, \bibinfo{person}{Andrea~Stevenson Won}, {and} \bibinfo{person}{Shiri Azenkot}.} \bibinfo{year}{2023}\natexlab{}.
\newblock \showarticletitle{“The Guide Has Your Back”: Exploring How Sighted Guides Can Enhance Accessibility in Social Virtual Reality for Blind and Low Vision People}. In \bibinfo{booktitle}{\emph{Proceedings of the 25th International ACM SIGACCESS Conference on Computers and Accessibility}}. \bibinfo{pages}{1--14}.
\newblock


\bibitem[Creed et~al\mbox{.}(2023)]%
        {creed2023inclusive}
\bibfield{author}{\bibinfo{person}{Chris Creed}, \bibinfo{person}{Maadh Al-Kalbani}, \bibinfo{person}{Arthur Theil}, \bibinfo{person}{Sayan Sarcar}, {and} \bibinfo{person}{Ian Williams}.} \bibinfo{year}{2023}\natexlab{}.
\newblock \showarticletitle{Inclusive augmented and virtual reality: A research agenda}.
\newblock \bibinfo{journal}{\emph{International Journal of Human--Computer Interaction}} (\bibinfo{year}{2023}), \bibinfo{pages}{1--20}.
\newblock


\bibitem[Franz et~al\mbox{.}(2023)]%
        {franz2023comparing}
\bibfield{author}{\bibinfo{person}{Rachel~L Franz}, \bibinfo{person}{Jinghan Yu}, {and} \bibinfo{person}{Jacob~O Wobbrock}.} \bibinfo{year}{2023}\natexlab{}.
\newblock \showarticletitle{Comparing Locomotion Techniques in Virtual Reality for People with Upper-Body Motor Impairments}. In \bibinfo{booktitle}{\emph{Proceedings of the 25th International ACM SIGACCESS Conference on Computers and Accessibility}}. \bibinfo{pages}{1--15}.
\newblock


\bibitem[Fung et~al\mbox{.}(2004)]%
        {fung2004locomotor}
\bibfield{author}{\bibinfo{person}{J Fung}, \bibinfo{person}{F Malouin}, \bibinfo{person}{BJ McFadyen}, \bibinfo{person}{F Comeau}, \bibinfo{person}{A Lamontagne}, \bibinfo{person}{S Chapdelaine}, \bibinfo{person}{C Beaudoin}, \bibinfo{person}{D Laurendeau}, \bibinfo{person}{L Hughey}, {and} \bibinfo{person}{CL Richards}.} \bibinfo{year}{2004}\natexlab{}.
\newblock \showarticletitle{Locomotor rehabilitation in a complex virtual environment}. In \bibinfo{booktitle}{\emph{The 26th Annual International Conference of the IEEE Engineering in Medicine and Biology Society}}, Vol.~\bibinfo{volume}{2}. IEEE, \bibinfo{pages}{4859--4861}.
\newblock


\bibitem[Garcia~Estrada and Prasolova-F{\o}rland(2022)]%
        {garcia2022running}
\bibfield{author}{\bibinfo{person}{Jose Garcia~Estrada} {and} \bibinfo{person}{Ekaterina Prasolova-F{\o}rland}.} \bibinfo{year}{2022}\natexlab{}.
\newblock \showarticletitle{Running an XR lab in the context of COVID-19 pandemic: Lessons learned from a Norwegian university}.
\newblock \bibinfo{journal}{\emph{Education and Information Technologies}} \bibinfo{volume}{27}, \bibinfo{number}{1} (\bibinfo{year}{2022}), \bibinfo{pages}{773--789}.
\newblock


\bibitem[Gerling et~al\mbox{.}(2020)]%
        {gerling2020virtual}
\bibfield{author}{\bibinfo{person}{Kathrin Gerling}, \bibinfo{person}{Patrick Dickinson}, \bibinfo{person}{Kieran Hicks}, \bibinfo{person}{Liam Mason}, \bibinfo{person}{Adalberto~L Simeone}, {and} \bibinfo{person}{Katta Spiel}.} \bibinfo{year}{2020}\natexlab{}.
\newblock \showarticletitle{Virtual reality games for people using wheelchairs}. In \bibinfo{booktitle}{\emph{Proceedings of the 2020 CHI Conference on Human Factors in Computing Systems}}. \bibinfo{pages}{1--11}.
\newblock


\bibitem[Gerling and Spiel(2021)]%
        {gerling2021critical}
\bibfield{author}{\bibinfo{person}{Kathrin Gerling} {and} \bibinfo{person}{Katta Spiel}.} \bibinfo{year}{2021}\natexlab{}.
\newblock \showarticletitle{A critical examination of virtual reality technology in the context of the minority body}. In \bibinfo{booktitle}{\emph{Proceedings of the 2021 CHI Conference on Human Factors in Computing Systems}}. \bibinfo{pages}{1--14}.
\newblock


\bibitem[Gerling et~al\mbox{.}(2014)]%
        {gerling2014effects}
\bibfield{author}{\bibinfo{person}{Kathrin~Maria Gerling}, \bibinfo{person}{Regan~L Mandryk}, \bibinfo{person}{Max~Valentin Birk}, \bibinfo{person}{Matthew Miller}, {and} \bibinfo{person}{Rita Orji}.} \bibinfo{year}{2014}\natexlab{}.
\newblock \showarticletitle{The effects of embodied persuasive games on player attitudes toward people using wheelchairs}. In \bibinfo{booktitle}{\emph{Proceedings of the SIGCHI Conference on Human Factors in Computing Systems}}. \bibinfo{pages}{3413--3422}.
\newblock


\bibitem[Gualano et~al\mbox{.}(2023)]%
        {gualano2023invisible}
\bibfield{author}{\bibinfo{person}{Ria~J Gualano}, \bibinfo{person}{Lucy Jiang}, \bibinfo{person}{Kexin Zhang}, \bibinfo{person}{Andrea~Stevenson Won}, {and} \bibinfo{person}{Shiri Azenkot}.} \bibinfo{year}{2023}\natexlab{}.
\newblock \showarticletitle{“Invisible Illness Is No Longer Invisible”: Making Social VR Avatars More Inclusive for Invisible Disability Representation}. In \bibinfo{booktitle}{\emph{Proceedings of the 25th International ACM SIGACCESS Conference on Computers and Accessibility}}. \bibinfo{pages}{1--4}.
\newblock


\bibitem[H.~Hoppe et~al\mbox{.}(2020)]%
        {h2020clevr}
\bibfield{author}{\bibinfo{person}{Adrian H.~Hoppe}, \bibinfo{person}{Julia~K Anken}, \bibinfo{person}{Thorsten Schwarz}, \bibinfo{person}{Rainer Stiefelhagen}, {and} \bibinfo{person}{Florian van~de Camp}.} \bibinfo{year}{2020}\natexlab{}.
\newblock \showarticletitle{CLEVR: A customizable interactive learning environment for users with low vision in virtual reality}. In \bibinfo{booktitle}{\emph{Proceedings of the 22nd International ACM SIGACCESS Conference on Computers and Accessibility}}. \bibinfo{pages}{1--4}.
\newblock


\bibitem[Hart and Staveland(1988)]%
        {hart1988development}
\bibfield{author}{\bibinfo{person}{Sandra~G Hart} {and} \bibinfo{person}{Lowell~E Staveland}.} \bibinfo{year}{1988}\natexlab{}.
\newblock \showarticletitle{Development of NASA-TLX (Task Load Index): Results of empirical and theoretical research}.
\newblock In \bibinfo{booktitle}{\emph{Advances in psychology}}. Vol.~\bibinfo{volume}{52}. \bibinfo{publisher}{Elsevier}, \bibinfo{pages}{139--183}.
\newblock


\bibitem[Hassan(2020)]%
        {hassan2020digitality}
\bibfield{author}{\bibinfo{person}{Robert Hassan}.} \bibinfo{year}{2020}\natexlab{}.
\newblock \showarticletitle{Digitality, virtual reality and the ‘empathy machine’}.
\newblock \bibinfo{journal}{\emph{Digital journalism}} \bibinfo{volume}{8}, \bibinfo{number}{2} (\bibinfo{year}{2020}), \bibinfo{pages}{195--212}.
\newblock


\bibitem[Herskovitz et~al\mbox{.}(2020)]%
        {herskovitz2020making}
\bibfield{author}{\bibinfo{person}{Jaylin Herskovitz}, \bibinfo{person}{Jason Wu}, \bibinfo{person}{Samuel White}, \bibinfo{person}{Amy Pavel}, \bibinfo{person}{Gabriel Reyes}, \bibinfo{person}{Anhong Guo}, {and} \bibinfo{person}{Jeffrey~P Bigham}.} \bibinfo{year}{2020}\natexlab{}.
\newblock \showarticletitle{Making mobile augmented reality applications accessible}. In \bibinfo{booktitle}{\emph{Proceedings of the 22nd International ACM SIGACCESS Conference on Computers and Accessibility}}. \bibinfo{pages}{1--14}.
\newblock


\bibitem[Ji et~al\mbox{.}(2022)]%
        {ji2022vrbubble}
\bibfield{author}{\bibinfo{person}{Tiger~F Ji}, \bibinfo{person}{Brianna Cochran}, {and} \bibinfo{person}{Yuhang Zhao}.} \bibinfo{year}{2022}\natexlab{}.
\newblock \showarticletitle{Vrbubble: Enhancing peripheral awareness of avatars for people with visual impairments in social virtual reality}. In \bibinfo{booktitle}{\emph{Proceedings of the 24th International ACM SIGACCESS Conference on Computers and Accessibility}}. \bibinfo{pages}{1--17}.
\newblock


\bibitem[Kennedy et~al\mbox{.}(1993)]%
        {kennedy1993simulator}
\bibfield{author}{\bibinfo{person}{Robert~S Kennedy}, \bibinfo{person}{Norman~E Lane}, \bibinfo{person}{Kevin~S Berbaum}, {and} \bibinfo{person}{Michael~G Lilienthal}.} \bibinfo{year}{1993}\natexlab{}.
\newblock \showarticletitle{Simulator sickness questionnaire: An enhanced method for quantifying simulator sickness}.
\newblock \bibinfo{journal}{\emph{The international journal of aviation psychology}} \bibinfo{volume}{3}, \bibinfo{number}{3} (\bibinfo{year}{1993}), \bibinfo{pages}{203--220}.
\newblock


\bibitem[Kim et~al\mbox{.}(1999)]%
        {kim1999new}
\bibfield{author}{\bibinfo{person}{Nam~Gyun Kim}, \bibinfo{person}{Choong~Ki Yoo}, {and} \bibinfo{person}{Jae~Joong Im}.} \bibinfo{year}{1999}\natexlab{}.
\newblock \showarticletitle{A new rehabilitation training system for postural balance control using virtual reality technology}.
\newblock \bibinfo{journal}{\emph{IEEE Transactions on Rehabilitation Engineering}} \bibinfo{volume}{7}, \bibinfo{number}{4} (\bibinfo{year}{1999}), \bibinfo{pages}{482--485}.
\newblock


\bibitem[Kroma et~al\mbox{.}(2022)]%
        {kroma2022reality}
\bibfield{author}{\bibinfo{person}{Assem Kroma}, \bibinfo{person}{Kristen Grinyer}, \bibinfo{person}{Anthony Scavarelli}, \bibinfo{person}{Elaheh Samimi}, \bibinfo{person}{Stanislav Kyian}, {and} \bibinfo{person}{Robert~J Teather}.} \bibinfo{year}{2022}\natexlab{}.
\newblock \showarticletitle{The reality of remote extended reality research: Practical case studies and taxonomy}.
\newblock \bibinfo{journal}{\emph{Frontiers in Computer Science}}  \bibinfo{volume}{4} (\bibinfo{year}{2022}), \bibinfo{pages}{954038}.
\newblock


\bibitem[L.~Franz et~al\mbox{.}(2021)]%
        {l2021nearmi}
\bibfield{author}{\bibinfo{person}{Rachel L.~Franz}, \bibinfo{person}{Sasa Junuzovic}, {and} \bibinfo{person}{Martez Mott}.} \bibinfo{year}{2021}\natexlab{}.
\newblock \showarticletitle{Nearmi: A framework for designing point of interest techniques for VR users with limited mobility}. In \bibinfo{booktitle}{\emph{Proceedings of the 23rd International ACM SIGACCESS Conference on Computers and Accessibility}}. \bibinfo{pages}{1--14}.
\newblock


\bibitem[Li et~al\mbox{.}(2022b)]%
        {li2022scoping}
\bibfield{author}{\bibinfo{person}{Yifan Li}, \bibinfo{person}{Kangsoo Kim}, \bibinfo{person}{Austin Erickson}, \bibinfo{person}{Nahal Norouzi}, \bibinfo{person}{Jonathan Jules}, \bibinfo{person}{Gerd Bruder}, {and} \bibinfo{person}{Gregory~F Welch}.} \bibinfo{year}{2022}\natexlab{b}.
\newblock \showarticletitle{A scoping review of assistance and therapy with head-mounted displays for people who are visually impaired}.
\newblock \bibinfo{journal}{\emph{ACM Transactions on Accessible Computing (TACCESS)}} \bibinfo{volume}{15}, \bibinfo{number}{3} (\bibinfo{year}{2022}), \bibinfo{pages}{1--28}.
\newblock


\bibitem[Li et~al\mbox{.}(2022a)]%
        {li2022soundvizvr}
\bibfield{author}{\bibinfo{person}{Ziming Li}, \bibinfo{person}{Shannon Connell}, \bibinfo{person}{Wendy Dannels}, {and} \bibinfo{person}{Roshan Peiris}.} \bibinfo{year}{2022}\natexlab{a}.
\newblock \showarticletitle{Soundvizvr: Sound indicators for accessible sounds in virtual reality for deaf or hard-of-hearing users}. In \bibinfo{booktitle}{\emph{Proceedings of the 24th International ACM SIGACCESS Conference on Computers and Accessibility}}. \bibinfo{pages}{1--13}.
\newblock


\bibitem[Mack et~al\mbox{.}(2021)]%
        {mack2021we}
\bibfield{author}{\bibinfo{person}{Kelly Mack}, \bibinfo{person}{Emma McDonnell}, \bibinfo{person}{Dhruv Jain}, \bibinfo{person}{Lucy Lu~Wang}, \bibinfo{person}{Jon E.~Froehlich}, {and} \bibinfo{person}{Leah Findlater}.} \bibinfo{year}{2021}\natexlab{}.
\newblock \showarticletitle{What do we mean by “accessibility research”? A literature survey of accessibility papers in CHI and ASSETS from 1994 to 2019}. In \bibinfo{booktitle}{\emph{Proceedings of the 2021 CHI Conference on Human Factors in Computing Systems}}. \bibinfo{pages}{1--18}.
\newblock


\bibitem[Mahmud et~al\mbox{.}(2023)]%
        {mahmud2023eyes}
\bibfield{author}{\bibinfo{person}{M~Rasel Mahmud}, \bibinfo{person}{Alberto Cordova}, {and} \bibinfo{person}{John Quarles}.} \bibinfo{year}{2023}\natexlab{}.
\newblock \showarticletitle{The Eyes Have It: Visual Feedback Methods to Make Walking in Immersive Virtual Reality More Accessible for People With Mobility Impairments While Utilizing Head-Mounted Displays}. In \bibinfo{booktitle}{\emph{Proceedings of the 25th International ACM SIGACCESS Conference on Computers and Accessibility}}. \bibinfo{pages}{1--10}.
\newblock


\bibitem[Maran et~al\mbox{.}(2022)]%
        {maran2022use}
\bibfield{author}{\bibinfo{person}{Patricia~Laura Maran}, \bibinfo{person}{Ramon Dani{\"e}ls}, {and} \bibinfo{person}{Karin Slegers}.} \bibinfo{year}{2022}\natexlab{}.
\newblock \showarticletitle{The use of extended reality (XR) for people with moderate to severe intellectual disabilities (ID): A scoping review}.
\newblock \bibinfo{journal}{\emph{Technology and Disability}} \bibinfo{volume}{34}, \bibinfo{number}{2} (\bibinfo{year}{2022}), \bibinfo{pages}{53--67}.
\newblock


\bibitem[Mathew et~al\mbox{.}(2022)]%
        {mathew2022access}
\bibfield{author}{\bibinfo{person}{Roshan Mathew}, \bibinfo{person}{Brian Mak}, {and} \bibinfo{person}{Wendy Dannels}.} \bibinfo{year}{2022}\natexlab{}.
\newblock \showarticletitle{Access on demand: real-time, multi-modal accessibility for the deaf and hard-of-hearing based on augmented reality}. In \bibinfo{booktitle}{\emph{Proceedings of the 24th International ACM SIGACCESS Conference on Computers and Accessibility}}. \bibinfo{pages}{1--6}.
\newblock


\bibitem[McGowan and Mcgregor(2023)]%
        {mcgowan2023investigation}
\bibfield{author}{\bibinfo{person}{John~Joseph McGowan} {and} \bibinfo{person}{Iain~Peter Mcgregor}.} \bibinfo{year}{2023}\natexlab{}.
\newblock \showarticletitle{Investigation into Stress Triggers in Autistic Adults for the Development of Technological Self-Interventions}. In \bibinfo{booktitle}{\emph{Proceedings of the 25th International ACM SIGACCESS Conference on Computers and Accessibility}}. \bibinfo{pages}{1--17}.
\newblock


\bibitem[Milgram and Kishino(1994)]%
        {milgram1994taxonomy}
\bibfield{author}{\bibinfo{person}{Paul Milgram} {and} \bibinfo{person}{Fumio Kishino}.} \bibinfo{year}{1994}\natexlab{}.
\newblock \showarticletitle{A taxonomy of mixed reality visual displays}.
\newblock \bibinfo{journal}{\emph{IEICE TRANSACTIONS on Information and Systems}} \bibinfo{volume}{77}, \bibinfo{number}{12} (\bibinfo{year}{1994}), \bibinfo{pages}{1321--1329}.
\newblock


\bibitem[Mott et~al\mbox{.}(2020)]%
        {mott2020just}
\bibfield{author}{\bibinfo{person}{Martez Mott}, \bibinfo{person}{John Tang}, \bibinfo{person}{Shaun Kane}, \bibinfo{person}{Edward Cutrell}, {and} \bibinfo{person}{Meredith Ringel~Morris}.} \bibinfo{year}{2020}\natexlab{}.
\newblock \showarticletitle{“i just went into it assuming that i wouldn't be able to have the full experience” understanding the accessibility of virtual reality for people with limited mobility}. In \bibinfo{booktitle}{\emph{Proceedings of the 22nd International ACM SIGACCESS Conference on Computers and Accessibility}}. \bibinfo{pages}{1--13}.
\newblock


\bibitem[Nabors et~al\mbox{.}(2020)]%
        {nabors2020scoping}
\bibfield{author}{\bibinfo{person}{Laura Nabors}, \bibinfo{person}{Julia Monnin}, {and} \bibinfo{person}{Solimar Jimenez}.} \bibinfo{year}{2020}\natexlab{}.
\newblock \showarticletitle{A scoping review of studies on virtual reality for individuals with intellectual disabilities}.
\newblock \bibinfo{journal}{\emph{Advances in Neurodevelopmental Disorders}} \bibinfo{volume}{4}, \bibinfo{number}{4} (\bibinfo{year}{2020}), \bibinfo{pages}{344--356}.
\newblock


\bibitem[Palaniappan et~al\mbox{.}(2019)]%
        {palaniappan2019identifying}
\bibfield{author}{\bibinfo{person}{Shanmugam~Muruga Palaniappan}, \bibinfo{person}{Ting Zhang}, {and} \bibinfo{person}{Bradley~S Duerstock}.} \bibinfo{year}{2019}\natexlab{}.
\newblock \showarticletitle{Identifying comfort areas in 3D space for persons with upper extremity mobility impairments using virtual reality}. In \bibinfo{booktitle}{\emph{Proceedings of the 21st International ACM SIGACCESS Conference on Computers and Accessibility}}. \bibinfo{pages}{495--499}.
\newblock


\bibitem[Pei et~al\mbox{.}(2023)]%
        {pei2023embodied}
\bibfield{author}{\bibinfo{person}{Siyou Pei}, \bibinfo{person}{Alexander Chen}, \bibinfo{person}{Chen Chen}, \bibinfo{person}{Franklin~Mingzhe Li}, \bibinfo{person}{Megan Fozzard}, \bibinfo{person}{Hao-Yun Chi}, \bibinfo{person}{Nadir Weibel}, \bibinfo{person}{Patrick Carrington}, {and} \bibinfo{person}{Yang Zhang}.} \bibinfo{year}{2023}\natexlab{}.
\newblock \showarticletitle{Embodied Exploration: Facilitating Remote Accessibility Assessment for Wheelchair Users with Virtual Reality}. In \bibinfo{booktitle}{\emph{Proceedings of the 25th International ACM SIGACCESS Conference on Computers and Accessibility}}. \bibinfo{pages}{1--17}.
\newblock


\bibitem[Pivik et~al\mbox{.}(2002)]%
        {pivik2002using}
\bibfield{author}{\bibinfo{person}{Jayne Pivik}, \bibinfo{person}{Joan McComas}, \bibinfo{person}{Ian MaCfarlane}, {and} \bibinfo{person}{Marc Laflamme}.} \bibinfo{year}{2002}\natexlab{}.
\newblock \showarticletitle{Using virtual reality to teach disability awareness}.
\newblock \bibinfo{journal}{\emph{Journal of Educational Computing Research}} \bibinfo{volume}{26}, \bibinfo{number}{2} (\bibinfo{year}{2002}), \bibinfo{pages}{203--218}.
\newblock


\bibitem[Powell and Myers(1995)]%
        {powell1995activities}
\bibfield{author}{\bibinfo{person}{Lynda~Elaine Powell} {and} \bibinfo{person}{Anita~M Myers}.} \bibinfo{year}{1995}\natexlab{}.
\newblock \showarticletitle{The activities-specific balance confidence (ABC) scale}.
\newblock \bibinfo{journal}{\emph{The Journals of Gerontology Series A: Biological Sciences and Medical Sciences}} \bibinfo{volume}{50}, \bibinfo{number}{1} (\bibinfo{year}{1995}), \bibinfo{pages}{M28--M34}.
\newblock


\bibitem[Quandt(2020)]%
        {quandt2020teaching}
\bibfield{author}{\bibinfo{person}{Lorna Quandt}.} \bibinfo{year}{2020}\natexlab{}.
\newblock \showarticletitle{Teaching ASL signs using signing avatars and immersive learning in virtual reality}. In \bibinfo{booktitle}{\emph{Proceedings of the 22nd International ACM SIGACCESS Conference on Computers and Accessibility}}. \bibinfo{pages}{1--4}.
\newblock


\bibitem[Riaz et~al\mbox{.}(2020)]%
        {riaz2020exploratory}
\bibfield{author}{\bibinfo{person}{Waleed Riaz}, \bibinfo{person}{Gulraiz Ali}, \bibinfo{person}{Momina Abid}, \bibinfo{person}{Izma~Naim Butt}, \bibinfo{person}{Anas Shahzad}, {and} \bibinfo{person}{Suleman Shahid}.} \bibinfo{year}{2020}\natexlab{}.
\newblock \showarticletitle{An Exploratory Study on Supporting Persons with Aphasia in Pakistan: Challenges and Opportunities}. In \bibinfo{booktitle}{\emph{Proceedings of the 22nd International ACM SIGACCESS Conference on Computers and Accessibility}}. \bibinfo{pages}{1--4}.
\newblock


\bibitem[Seol et~al\mbox{.}(2017)]%
        {seol2017drop}
\bibfield{author}{\bibinfo{person}{Eunbi Seol}, \bibinfo{person}{Seulki Min}, \bibinfo{person}{Sungho Seo}, \bibinfo{person}{Seoyeon Jung}, \bibinfo{person}{Youngil Lee}, \bibinfo{person}{Jaedong Lee}, \bibinfo{person}{Gerard Kim}, \bibinfo{person}{Chungyean Cho}, \bibinfo{person}{Seungmoo Lee}, \bibinfo{person}{Chul-Hyun Cho}, {et~al\mbox{.}}} \bibinfo{year}{2017}\natexlab{}.
\newblock \showarticletitle{" Drop the beat" virtual reality based mindfulness and cognitive behavioral therapy for panic disorder---a pilot study}. In \bibinfo{booktitle}{\emph{Proceedings of the 23rd acm symposium on virtual reality software and technology}}. \bibinfo{pages}{1--3}.
\newblock


\bibitem[Shew(2020)]%
        {shew2020ableism}
\bibfield{author}{\bibinfo{person}{Ashley Shew}.} \bibinfo{year}{2020}\natexlab{}.
\newblock \showarticletitle{Ableism, technoableism, and future AI}.
\newblock \bibinfo{journal}{\emph{IEEE Technology and Society Magazine}} \bibinfo{volume}{39}, \bibinfo{number}{1} (\bibinfo{year}{2020}), \bibinfo{pages}{40--85}.
\newblock


\bibitem[Sidarto et~al\mbox{.}({[n.\,d.]})]%
        {sidartoaccessibility}
\bibfield{author}{\bibinfo{person}{Lauren Sidarto}, \bibinfo{person}{Kayla~A Monnette}, {and} \bibinfo{person}{Cheuk~Wah Chim}.} \bibinfo{year}{[n.\,d.]}\natexlab{}.
\newblock \showarticletitle{Accessibility in Extended Reality}.
\newblock  (\bibinfo{year}{[n.\,d.]}).
\newblock


\bibitem[Siu et~al\mbox{.}(2020)]%
        {siu2020virtual}
\bibfield{author}{\bibinfo{person}{Alexa~F Siu}, \bibinfo{person}{Mike Sinclair}, \bibinfo{person}{Robert Kovacs}, \bibinfo{person}{Eyal Ofek}, \bibinfo{person}{Christian Holz}, {and} \bibinfo{person}{Edward Cutrell}.} \bibinfo{year}{2020}\natexlab{}.
\newblock \showarticletitle{Virtual reality without vision: A haptic and auditory white cane to navigate complex virtual worlds}. In \bibinfo{booktitle}{\emph{Proceedings of the 2020 CHI conference on human factors in computing systems}}. \bibinfo{pages}{1--13}.
\newblock


\bibitem[Su et~al\mbox{.}(2023)]%
        {su2023demonstration}
\bibfield{author}{\bibinfo{person}{Xia Su}, \bibinfo{person}{Kaiming Cheng}, \bibinfo{person}{Han Zhang}, \bibinfo{person}{Jaewook Lee}, \bibinfo{person}{Wyatt Olson}, {and} \bibinfo{person}{Jon~E Froehlich}.} \bibinfo{year}{2023}\natexlab{}.
\newblock \showarticletitle{A Demonstration of RASSAR: Room Accessibility and Safety Scanning in Augmented Reality}. In \bibinfo{booktitle}{\emph{Proceedings of the 25th International ACM SIGACCESS Conference on Computers and Accessibility}}. \bibinfo{pages}{1--4}.
\newblock


\bibitem[Sutton et~al\mbox{.}(2019)]%
        {sutton2019meeting}
\bibfield{author}{\bibinfo{person}{Anthea Sutton}, \bibinfo{person}{Mark Clowes}, \bibinfo{person}{Louise Preston}, {and} \bibinfo{person}{Andrew Booth}.} \bibinfo{year}{2019}\natexlab{}.
\newblock \showarticletitle{Meeting the review family: exploring review types and associated information retrieval requirements}.
\newblock \bibinfo{journal}{\emph{Health Information \& Libraries Journal}} \bibinfo{volume}{36}, \bibinfo{number}{3} (\bibinfo{year}{2019}), \bibinfo{pages}{202--222}.
\newblock


\bibitem[Sveistrup(2004)]%
        {sveistrup2004motor}
\bibfield{author}{\bibinfo{person}{Heidi Sveistrup}.} \bibinfo{year}{2004}\natexlab{}.
\newblock \showarticletitle{Motor rehabilitation using virtual reality}.
\newblock \bibinfo{journal}{\emph{Journal of neuroengineering and rehabilitation}}  \bibinfo{volume}{1} (\bibinfo{year}{2004}), \bibinfo{pages}{1--8}.
\newblock


\bibitem[Thevin et~al\mbox{.}(2020)]%
        {thevin2020x}
\bibfield{author}{\bibinfo{person}{Lauren Thevin}, \bibinfo{person}{Carine Briant}, {and} \bibinfo{person}{Anke~M Brock}.} \bibinfo{year}{2020}\natexlab{}.
\newblock \showarticletitle{X-road: virtual reality glasses for orientation and mobility training of people with visual impairments}.
\newblock \bibinfo{journal}{\emph{ACM Transactions on Accessible Computing (TACCESS)}} \bibinfo{volume}{13}, \bibinfo{number}{2} (\bibinfo{year}{2020}), \bibinfo{pages}{1--47}.
\newblock


\bibitem[Tricco et~al\mbox{.}(2018)]%
        {tricco2018prisma}
\bibfield{author}{\bibinfo{person}{Andrea~C Tricco}, \bibinfo{person}{Erin Lillie}, \bibinfo{person}{Wasifa Zarin}, \bibinfo{person}{Kelly~K O'Brien}, \bibinfo{person}{Heather Colquhoun}, \bibinfo{person}{Danielle Levac}, \bibinfo{person}{David Moher}, \bibinfo{person}{Micah~DJ Peters}, \bibinfo{person}{Tanya Horsley}, \bibinfo{person}{Laura Weeks}, {et~al\mbox{.}}} \bibinfo{year}{2018}\natexlab{}.
\newblock \showarticletitle{PRISMA extension for scoping reviews (PRISMA-ScR): checklist and explanation}.
\newblock \bibinfo{journal}{\emph{Annals of internal medicine}} \bibinfo{volume}{169}, \bibinfo{number}{7} (\bibinfo{year}{2018}), \bibinfo{pages}{467--473}.
\newblock


\bibitem[Troncoso~Aldas et~al\mbox{.}(2020)]%
        {troncoso2020aiguide}
\bibfield{author}{\bibinfo{person}{Nelson~Daniel Troncoso~Aldas}, \bibinfo{person}{Sooyeon Lee}, \bibinfo{person}{Chonghan Lee}, \bibinfo{person}{Mary~Beth Rosson}, \bibinfo{person}{John~M Carroll}, {and} \bibinfo{person}{Vijaykrishnan Narayanan}.} \bibinfo{year}{2020}\natexlab{}.
\newblock \showarticletitle{AIGuide: An augmented reality hand guidance application for people with visual impairments}. In \bibinfo{booktitle}{\emph{Proceedings of the 22nd International ACM SIGACCESS Conference on Computers and Accessibility}}. \bibinfo{pages}{1--13}.
\newblock


\bibitem[Ventura et~al\mbox{.}(2020)]%
        {ventura2020virtual}
\bibfield{author}{\bibinfo{person}{Sara Ventura}, \bibinfo{person}{Laura Badenes-Ribera}, \bibinfo{person}{Rocio Herrero}, \bibinfo{person}{Ausias Cebolla}, \bibinfo{person}{Laura Galiana}, {and} \bibinfo{person}{Rosa Ba{\~n}os}.} \bibinfo{year}{2020}\natexlab{}.
\newblock \showarticletitle{Virtual reality as a medium to elicit empathy: A meta-analysis}.
\newblock \bibinfo{journal}{\emph{Cyberpsychology, Behavior, and Social Networking}} \bibinfo{volume}{23}, \bibinfo{number}{10} (\bibinfo{year}{2020}), \bibinfo{pages}{667--676}.
\newblock


\bibitem[Wentzel(2023)]%
        {wentzel2023bring}
\bibfield{author}{\bibinfo{person}{Johann Wentzel}.} \bibinfo{year}{2023}\natexlab{}.
\newblock \showarticletitle{Bring-Your-Own Input: Context-Aware Multi-Modal Input for More Accessible Virtual Reality}. In \bibinfo{booktitle}{\emph{Extended Abstracts of the 2023 CHI Conference on Human Factors in Computing Systems}}. \bibinfo{pages}{1--5}.
\newblock


\bibitem[Wilson et~al\mbox{.}(1997)]%
        {wilson1997virtual}
\bibfield{author}{\bibinfo{person}{Paul~N Wilson}, \bibinfo{person}{Nigel Foreman}, {and} \bibinfo{person}{Dana{\"e} Stanton}.} \bibinfo{year}{1997}\natexlab{}.
\newblock \showarticletitle{Virtual reality, disability and rehabilitation}.
\newblock \bibinfo{journal}{\emph{Disability and rehabilitation}} \bibinfo{volume}{19}, \bibinfo{number}{6} (\bibinfo{year}{1997}), \bibinfo{pages}{213--220}.
\newblock


\bibitem[Yoon et~al\mbox{.}(2019)]%
        {yoon2019leveraging}
\bibfield{author}{\bibinfo{person}{Chris Yoon}, \bibinfo{person}{Ryan Louie}, \bibinfo{person}{Jeremy Ryan}, \bibinfo{person}{MinhKhang Vu}, \bibinfo{person}{Hyegi Bang}, \bibinfo{person}{William Derksen}, {and} \bibinfo{person}{Paul Ruvolo}.} \bibinfo{year}{2019}\natexlab{}.
\newblock \showarticletitle{Leveraging augmented reality to create apps for people with visual disabilities: A case study in indoor navigation}. In \bibinfo{booktitle}{\emph{Proceedings of the 21st International ACM SIGACCESS Conference on Computers and Accessibility}}. \bibinfo{pages}{210--221}.
\newblock


\bibitem[Zhang et~al\mbox{.}(2022)]%
        {zhang2022s}
\bibfield{author}{\bibinfo{person}{Kexin Zhang}, \bibinfo{person}{Elmira Deldari}, \bibinfo{person}{Zhicong Lu}, \bibinfo{person}{Yaxing Yao}, {and} \bibinfo{person}{Yuhang Zhao}.} \bibinfo{year}{2022}\natexlab{}.
\newblock \showarticletitle{“it’s just part of me:” understanding avatar diversity and self-presentation of people with disabilities in social virtual reality}. In \bibinfo{booktitle}{\emph{Proceedings of the 24th international ACM SIGACCESS conference on computers and accessibility}}. \bibinfo{pages}{1--16}.
\newblock


\bibitem[Zhang et~al\mbox{.}(2023)]%
        {zhang2023diary}
\bibfield{author}{\bibinfo{person}{Kexin Zhang}, \bibinfo{person}{Elmira Deldari}, \bibinfo{person}{Yaxing Yao}, {and} \bibinfo{person}{Yuhang Zhao}.} \bibinfo{year}{2023}\natexlab{}.
\newblock \showarticletitle{A Diary Study in Social Virtual Reality: Impact of Avatars with Disability Signifiers on the Social Experiences of People with Disabilities}. In \bibinfo{booktitle}{\emph{Proceedings of the 25th International ACM SIGACCESS Conference on Computers and Accessibility}}. \bibinfo{pages}{1--17}.
\newblock


\end{thebibliography}

\end{document}